# Photomeson-threshold origin of the TeV break in astrophysical neutrinos: Spectrum, flavor, resolved sources, and the ultrahigh-energy landscape

Vernon Barger[*]

*Department of Physics, University of Wisconsin–Madison, Madison, Wisconsin 53706, USA*

IceCube observes a break near 30 TeV in the diffuse astrophysical neutrino spectrum. We interpret it as the photomeson ($p\gamma$) threshold. Accelerated protons interact with the soft x rays of accreting black holes through the $\Delta(1232)$ resonance. One kinematic relation sets the break energy from the target field's lowest photon energy, the seed scale $\varepsilon_0$. The measured break fixes $\varepsilon_0 \simeq 0.13$ keV and supplies the low-energy suppression no sum of falling power laws can produce. Building on the $\Delta(1232)$ identification of Khatee Zathul, Fang, Halzen, and Hooper and the single-source-class photo-pion fit of Winter, Fiorillo, and Buson, we give closed-form relations that invert the four broken-power-law fit parameters to the source physics with no remaining freedom, namely, $\gamma_2 = s$ and $\gamma_1 = s - (\alpha_x - 1)$ ($s$, $\alpha_x$ the proton and target photon indices), and $E_{\rm br} \propto [(1+\bar{z})\,\varepsilon_0]^{-1}$. The break width then measures the population's uniformity. The total smearing is bounded to $\sigma_s \leqslant 0.235$ dex (68% confidence level), no broader than redshift alone must make it, and the seed scale is the same across the sources to within a factor $\lesssim 1.5$, a constraint mean-spectrum fits do not provide. The model also predicts that below-break neutrinos come on average from higher redshifts than those above. The threshold also fixes the flavor and neutrino-antineutrino composition, including the energy-dependent $\bar{\nu}_e$ fraction and the Glashow-resonance discriminator computed in detail by Winter *et al.*, matching current flavor data. Inverted source by source, the relation gives each resolved IceCube source its target photon scale, from the keV coronae of NGC 1068 and PKS 1424+240 to the $\sim$20 eV jet of TXS 0506+056. That two-decade spread means they cannot supply the homogeneous break. It requires a separate, fainter population. KM3-230213A points to a warm-dust target, $\varepsilon_0 \simeq 0.02$ eV, unless cosmogenic.

## I. INTRODUCTION

More than a decade after IceCube established a diffuse flux of high-energy astrophysical neutrinos [1], its spectrum has resolved into a distinct feature, a break near 30 TeV, below which the spectrum is hard and above which it is soft [2]. Two independent analyses, a medium-energy starting events (MESE) sample and a combined track-cascade fit (CF), reject a single power law over 5 TeV–10 PeV at $>4\sigma$ and prefer the broken power law (BPL) of Table I, with $\gamma_1 \simeq 1.3$–$1.7$ below the break, $\gamma_2 \simeq 2.7$–$2.8$ above it, and $E_{\rm br} \simeq 25$–$33$ TeV. The question this paper addresses is what physics puts the break there, and what follows once that physics is identified.

[*]Contact author: barger@pheno.wisc.edu

TABLE I. IceCube's broken-power-law fit [2] and the source parameters it implies through the threshold relations ($\bar{z} = 1$). $\phi_0$ in $10^{-18}$ GeV$^{-1}$ cm$^{-2}$ s$^{-1}$ sr$^{-1}$ at 100 TeV, all flavors.

| | MESE | CF |
|---|---|---|
| $\phi_0$ | $2.28^{+0.22}_{-0.20}$ | $1.77^{+0.15}_{-0.11}$ |
| $\gamma_1$ | $1.72^{+0.26}_{-0.35}$ | $1.31^{+0.50}_{-1.21}$ |
| $\gamma_2$ | $2.84^{+0.11}_{-0.09}$ | $2.74^{+0.06}_{-0.07}$ |
| $\log_{10}(E_{\rm br}/\mathrm{GeV})$ | $4.52^{+0.11}_{-0.09}$ | $4.39^{+0.09}_{-0.08}$ |
| $s = \gamma_2$ | $2.84 \pm 0.10$ | $2.74 \pm 0.07$ |
| $\alpha_x = 1 + (\gamma_2 - \gamma_1)$ | $2.12 \pm 0.32$ | $2.43 \pm 0.86$ |
| $\varepsilon_0$ $[(1+\bar{z})/2]$ [eV] | $\simeq$130 | $\simeq$175 |

The answer we develop is the threshold of photomeson ($p\gamma$) production. Protons accelerated near an accreting supermassive black hole interact with the soft x-ray photons of its corona through the $\Delta(1232)$ resonance. The resonance pairs each proton energy with a definite photon energy, and higher-energy protons pair with lower-energy photons. The target photon spectrum has a lowest energy, which we call the *seed scale* $\varepsilon_0$. In practice, it is set by the temperature of the soft-x-ray "warm corona" of the accretion flow. The seed scale fixes the proton energy at which the interaction probability stops growing, and that proton energy maps to the neutrino break. One kinematic relation therefore connects the measured break energy to the seed scale, and the measured value returns $\varepsilon_0 \simeq 0.13$ keV.

This interpretation was recently developed by Khatee Zathul *et al.* [3], who reproduced the diffuse spectrum and relieved its tension with the isotropic $\gamma$-ray background, using protons of index $\simeq$3.1 on $\sim$0.3 keV x rays through the $\Delta(1232)$ resonance, and by Winter *et al.* [4], whose single-source-class photo-pion fit on a thermal target scans the proton index, maximal energy, target temperature, and magnetic field. We adopt that flux description and develop consequences it leaves open.

First, closed-form threshold relations invert the four BPL fit parameters to the source physics with no remaining freedom (Fig. 1). Second, the width of the break bounds how much the seed scale can vary across the source population, a constraint the mean-spectrum fits do not provide. Third, the same population model predicts which source redshifts contribute at each observed energy. The typical redshift is higher below the break than above it, so the break also marks a change in the distances of the contributing sources.

Fourth, the threshold fixes the flavor and neutrino-antineutrino composition. The detailed six-species decomposition, with its energy-dependent $\nu/\bar{\nu}$ structure and energy-independent flavor sum, was computed by Winter *et al.* [4], building on the Glashow resonance [5] as used in the source-diagnostic framework of Ref. [6]. We recover its main features analytically and add the propagation through the Earth. There, the Mikheyev-Smirnov-Wolfenstein (MSW) matter effect [7–9], evaluated with the Preliminary Reference Earth Model (PREM) [10] density profile, changes the flavor conversion probabilities by less than $5\times10^{-4}$ everywhere in the band, so the Earth leaves the flavor content unchanged. Absorption, however, confines detectable Glashow-resonance $\bar{\nu}_e$ to directions above the horizon. Fifth, the same break relation, inverted source by source, assigns each resolved IceCube source a target photon scale. The resolved sources turn out to span very different targets from the homogeneous break population. The ultrahigh-energy event KM3-230213A is placed in the same framework.

The paper is organized as follows. Section II shows that the break cannot be assembled from falling power laws and requires intrinsic suppression, for which the photomeson threshold is the generic mechanism. Section III presents the threshold relations and the resulting fit. Sections IV and V develop the population uniformity bound and the predicted source-redshift distribution. Section VI gives the flavor and antineutrino structure. Sections VII and VIII treat KM3-230213A and the resolved sources, and Sec. IX collects the observational consequences.

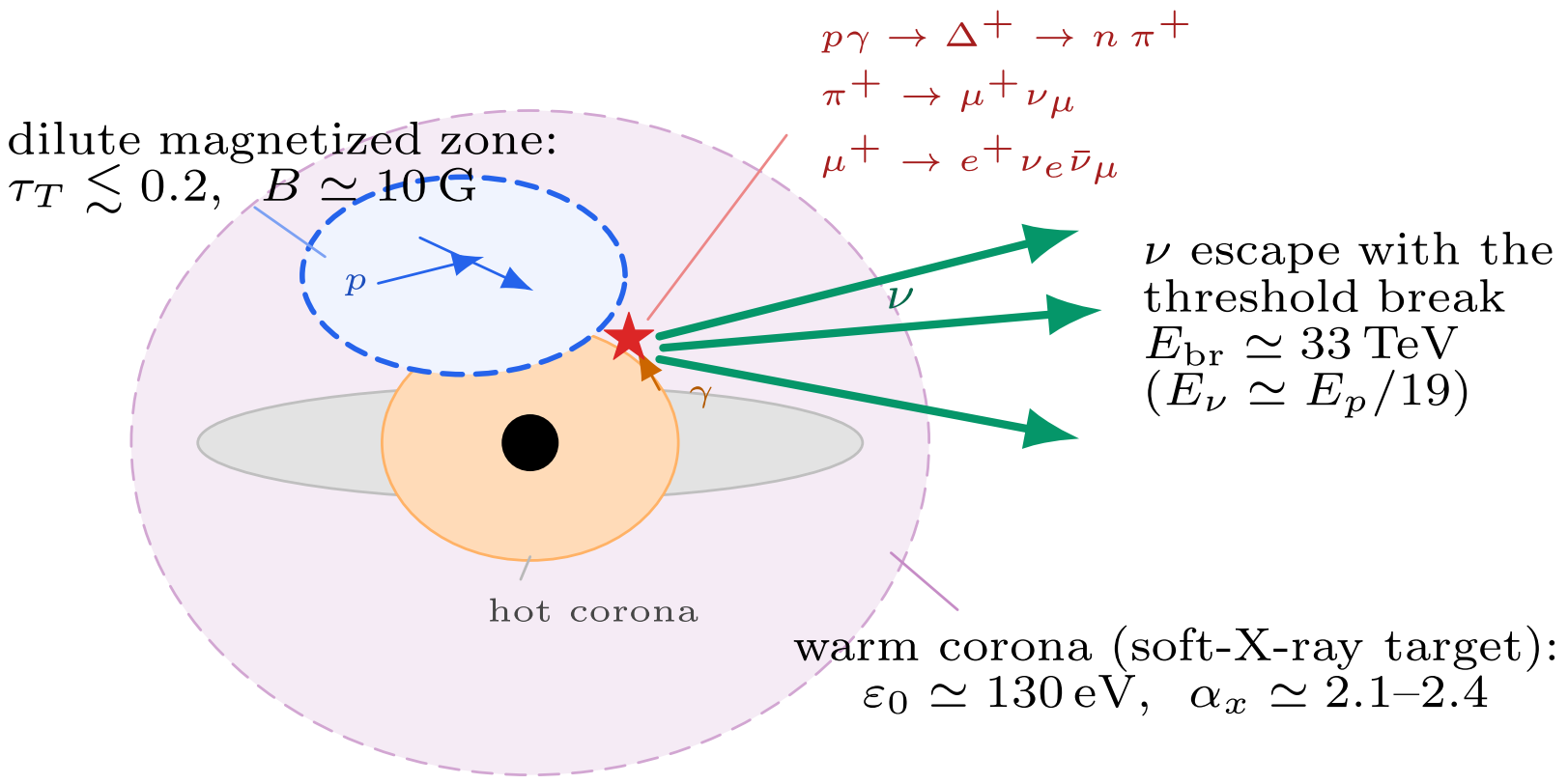

FIG. 1. Where the break forms, annotated with the quantities this work measures or bounds. Protons accelerated in a dilute magnetized zone (Thomson depth $\tau_T \lesssim 0.2$, field $B \simeq 10$ G) are embedded in the warm-corona soft-x-ray field ($\varepsilon_0 \simeq 130$ eV, target index $\alpha_x \simeq 2.1$–$2.4$, Table I) that surrounds the Comptonizing hot corona. Photomeson production at the $\Delta(1232)$ resonance imprints the threshold break. The charged-pion branch $\Delta^+ \to n\pi^+$, followed by $\pi^+ \to \mu^+\nu_\mu$, produces the neutrinos, which escape carrying the break at $E_{\rm br} \simeq 33$ TeV ($E_\nu \simeq E_p/19$).

## II. THE BREAK REQUIRES INTRINSIC SUPPRESSION

A spectrum summed from falling power laws can sag into a valley but cannot rise into a peak. IceCube's diffuse astrophysical neutrino flux rises into one. In $E^2\Phi$, the BPL of Table I rises below the break (slope $2 - \gamma_1 \simeq +0.3$) and falls above it ($2 - \gamma_2 \simeq -0.8$). A sum of unbroken falling power laws is convex in the $\log_{10} E$–$\log_{10}(E^2\Phi)$ plane and can produce valleys, never peaks, so the population dominating near the break must be intrinsically suppressed below $\sim$30 TeV.

Among the candidate mechanisms (a kinematic threshold [3,4], a minimum proton energy near a PeV that no acceleration model singles out, peaked dark-matter injection [11,12] constrained by isotropy and the absence of halo anisotropy, or low-energy neutrino absorption through new interactions [13,14] with no standard-model realization), only the photomeson threshold is generic. Every $p\gamma$ source has one.

## III. THRESHOLD MODEL

### A. Threshold relations

Protons with $dN_p/dE_p \propto E_p^{-s}$ interact with an isotropic field $dn/d\varepsilon \propto \varepsilon^{-\alpha_x}$ truncated at the seed scale $\varepsilon_0$, the lowest photon energy in the target field. The $\Delta(1232)$ resonance pairs a proton of energy $E_p$ with photons of resonant energy $\varepsilon_r \simeq (m_\Delta^2 - m_p^2)/4E_p \simeq 0.16\,\mathrm{GeV}^2/E_p$ (the head-on peak condition), so higher-energy protons sample lower-energy photons. Averaging over the isotropic target shifts this by an $O(1)$ factor absorbed into the effective seed scale $\varepsilon_0$. The photomeson optical depth $\tau_{p\gamma}$, the mean number of $p\gamma$ interactions a proton undergoes, then tracks the target field at $\varepsilon_r$ and rises as $\tau_{p\gamma} \propto E_p^{\alpha_x - 1}$ while $\varepsilon_r > \varepsilon_0$, saturating on the multipion plateau once $2\gamma_p\varepsilon_0 > \varepsilon'_\Delta = (m_\Delta^2 - m_p^2)/2m_p$, with $\gamma_p = E_p/m_p$ the proton Lorentz factor (Fig. 2).

Figure 2 is schematic. The Appendix computes the single-source spectrum with the full $p\gamma$ cross section [the $\Delta(1232)$ Breit-Wigner plus the multipion plateau], the isotropic-target angle average, and the decay kinematics of the secondaries. Three lessons carry forward. The asymptotic slopes that fix $\gamma_1$ and $\gamma_2$ survive intact, since a fixed decay kernel preserves power laws. The resonance width and the angle average barely round the observable break. Their effect is absorbed into the break position. The rounding that remains is dominated by the spread of neutrino energies in the decay chain $\pi^+ \to \mu^+\nu_\mu$, $\mu^+ \to e^+\nu_e\bar{\nu}_\mu$, and its size depends on the target. The thermal target gives an effective single-source width $\sigma_{\rm thr} \simeq 0.14$ dex, while a sharply truncated power-law target would give $\simeq$0.3 dex (Sec. IV).

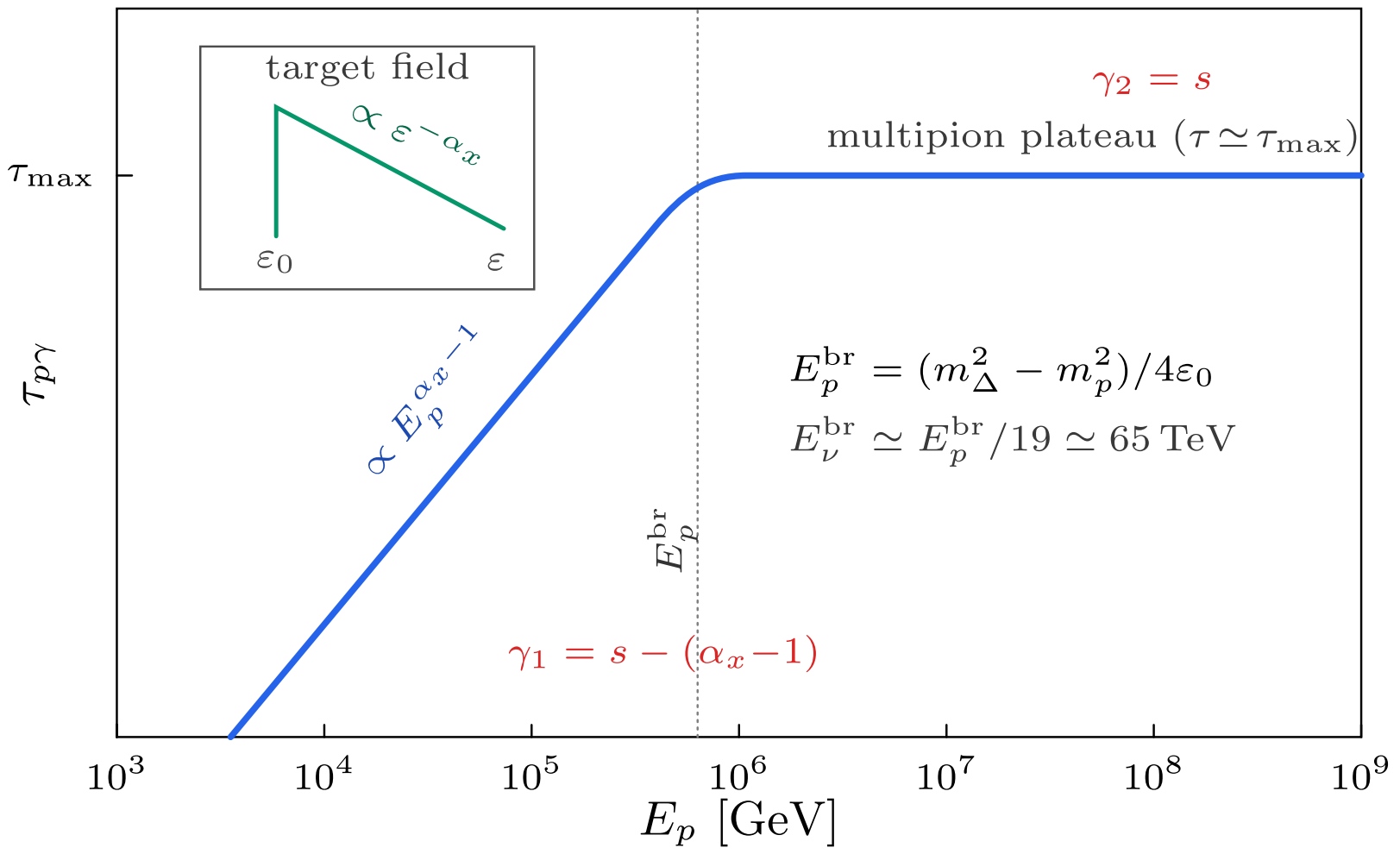

FIG. 2. The break as a turnover in photomeson optical depth (schematic, and the vertical normalization is arbitrary). For a proton spectrum $\propto E_p^{-s}$ on the truncated target field of the inset (photon index $\alpha_x$, low-energy cutoff $\varepsilon_0$), $\tau_{p\gamma}$ rises as $E_p^{\alpha_x-1}$ while the resonant photon energy stays above the cutoff ($\varepsilon_r > \varepsilon_0$) and saturates on the multipion plateau once $2\gamma_p\varepsilon_0 > \varepsilon'_\Delta$, turning over at $E_p^{\rm br} = (m_\Delta^2 - m_p^2)/4\varepsilon_0$ (source-frame $E_\nu^{\rm br} \simeq E_p^{\rm br}/19 \simeq$ 65 TeV, diluting to the observed $\simeq$33 TeV through $1+\bar{z}$). Because the neutrino flux is the proton flux times this depth, $\Phi_\nu \propto E^{-s}\tau_{p\gamma}$, the rising branch hardens the spectrum to $\gamma_1 = s - (\alpha_x - 1)$ below the break while the plateau returns $\gamma_2 = s$ above it, recovering Eqs. (1) and (2). A realistic calculation leaves the asymptotic slopes, and hence $\gamma_1$ and $\gamma_2$, unaffected. The observable rounding of the break is dominated not by the cross section but by the decay kinematics of the secondaries, and is quantified in the Appendix.

With $E_\nu \simeq E_p/19$ and redshift dilution,

$$\gamma_2 = s, \qquad \gamma_1 = s - (\alpha_x - 1), \tag{1}$$

$$E_{\rm br} \simeq \frac{\kappa_\pi}{16}\frac{m_\Delta^2 - m_p^2}{(1+\bar{z})\,\varepsilon_0} \simeq 33\,{\rm TeV}\left(\frac{\varepsilon_0}{0.13\,{\rm keV}}\right)^{-1}\left(\frac{1+\bar{z}}{2}\right)^{-1}. \tag{2}$$

A concrete flux follows from the thermal-target photo-pion calculation of Winter *et al.* [4], built on the Hümmer *et al.* [15] parametrization of the SOPHIA photohadronic code of Mücke *et al.* [16]. Protons $N_p \propto E_p^{-\alpha_p}\exp(-E_p/E_{p,\max})$ on a blackbody of temperature $T$ interact once in the optically thin regime. The threshold imprints the break, and multipion production carries the high-energy neutrinos along the proton spectrum, $\gamma_2 \simeq \alpha_p$, up to $E_{\nu,\max} \simeq 0.05\,E_{p,\max}$. The source magnetic field $B$ enters because the secondary pions, muons, and kaons can lose energy to synchrotron radiation before they decay. A strong field would steepen the neutrino spectrum above a $B$-dependent energy and alter the flavor mix, which is why Winter *et al.* scan it. For the coronal benchmark $B \simeq 10$ G, these cooling breaks lie far above the band (Sec. VI), so the decay spectra here are unmodified. Heavy nuclei in the beam would photodisintegrate on this field below the photomeson threshold [the giant-dipole resonance lies roughly an order of magnitude lower in per-nucleon energy than $\Delta(1232)$], so protons are the natural beam. The thermal scale sets $\varepsilon_0$ in Eq. (2), with $\varepsilon_0 \simeq k_B T$ to within an $O(1)$ factor and the soft-x-ray field's mean photon energy $\simeq 3k_B T \simeq 0.3$ keV. The best-fit $k_B T \simeq 0.12$ keV [3,4] then reproduces the measured break (Table I). Figure 3 shows the soft-power-law best fit (scenario A of Ref. [4], $\alpha_p = 3$, $k_B T = 0.12$ keV) and a hard variant with a proton cutoff (scenario B, $\alpha_p = 2$, $E_{p,\max} = 0.56$ PeV) against the IceCube BPL best fit and its 68% band. The soft scenario tracks the break and the high-energy spectrum, whereas the cutoff variant terminates near the break and supplies essentially no flux toward the Glashow resonance. A second, subdominant component is needed to carry the highest-energy flux. We model it as a population of sources optically thin ("transparent") to their own GeV-TeV $\gamma$ rays, in contrast to the $\gamma$-absorbing corona. Their neutrinos come from proton-gas ($pp$) collisions, trace the proton spectrum with no photomeson threshold, and carry equal $\nu$ and $\bar{\nu}$ of every flavor. This is the gray dashed curve of Fig. 3, normalized to the measured high-energy flux. Its role in the flavor structure is developed in Sec. VI. The soft scenario-A corona tail reaches the same all-flavor bins, but the two differ in $\bar{\nu}_e$ at the Glashow resonance, where the transparent component supplies the order-unity rate.

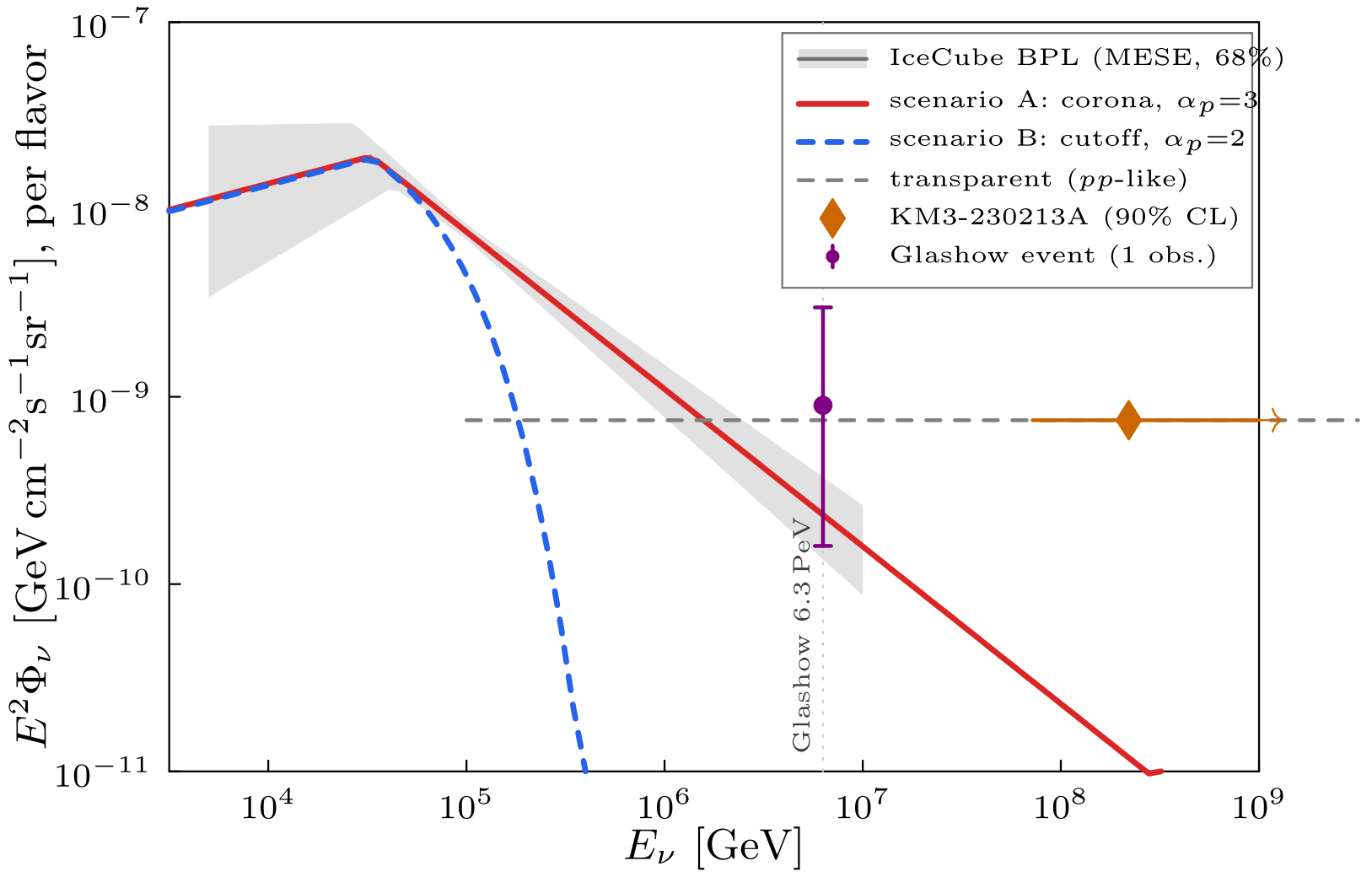

FIG. 3. The thermal-target photomeson spectrum reproduces IceCube's break. The broken-power-law best fit with its 68% band (MESE fit, Table I) is shown with the single-source-class description of Winter *et al.* [4] that builds on Khatee Zathul *et al.* [3]. The solid curve is the soft-power-law best fit (scenario A, $\alpha_p = 3$, $k_B T = 0.12$ keV), hard below the threshold break and following the proton spectrum, $E^{2-\alpha_p}$, above it through multipion production. The blue dashed curve is a hard variant with a proton cutoff (scenario B, $\alpha_p = 2$, $E_{p,\max} = 0.56$ PeV). The gray dashed curve is a second component from sources optically thin ("transparent") to their own $\gamma$ rays, whose neutrinos come from proton-gas ($pp$) collisions and carry no threshold (Sec. VI). It is normalized to the measured high-energy flux and extended to $E_{p,\max} \sim$ few EeV. The same curve passes the Glashow energy (6.3 PeV, vertical marker), where its $\bar{\nu}_e$ gives an order-unity rate, and KM3-230213A at 220 PeV (orange diamond, at the joint IceCube+KM3 normalization $E^2\Phi^{1f} = 7.5\times10^{-10}$ GeV cm$^{-2}$ s$^{-1}$ sr$^{-1}$ single flavor, with the 90% energy interval). At this IceCube-consistent level, KM3-230213A is a $\sim 2\sigma$ upward fluctuation rather than a reproduced event. The violet point at 6.3 PeV is the single observed Glashow-resonance event [17], placed at the per-flavor flux reproducing the observed rate under the democratic $\bar{\nu}_e$ fraction. Its bar is the 68% Poisson interval for one count, consistent with the transparent normalization to which the order-unity rate is matched. The flat continuation sits at the IceCube intermediate limits, where the mild tension with the EHE search arises. A warm-dust peak at 220 PeV would relax it. The corona shapes follow the benchmark parameters. Only their normalization is fit to the IceCube flux.

### B. Inversion to the source physics

We adopt the MESE fit as the baseline throughout. Its starting-event sample reaches below 10 TeV and localizes the break most tightly, and the CF values are consistent and listed alongside in Table I. The four measured BPL parameters thus invert, with no remaining freedom, to the source physics in Table I. The result is a soft, stochastically accelerated proton spectrum, a target photon index in the warm-Comptonization range, and a seed scale in the extreme-ultraviolet/soft-x-ray band of accreting supermassive black holes (Fig. 1).

The hard $\gamma_1$ is not an accelerator property but the threshold suppression imprinted on a soft spectrum, and the same photons absorb GeV-TeV $\gamma$ rays ($\tau_{\gamma\gamma} \sim 10^3 \tau_{p\gamma}$), so the sources are $\gamma$-ray dim [18–20], consistent with the *Fermi* background.

### C. Fit to the measured flux

The analytic shape fixed entirely by Eqs. (1) and (2), with only the normalization floated, reproduces the same segmented flux [21] ($\chi^2/\mathrm{dof} = 5.9/9$ and $8.9/9$ for the two samples), an equivalent description to the photo-pion calculation of Fig. 3.

Beyond the mean spectrum, fitting the same shape broadened by a Gaussian bounds the total break smearing to $\sigma_s \leqslant 0.235$ dex (68% confidence level, CL), which constrains the dispersion of the seed scale across the source population (Sec. IV).

## IV. POPULATION UNIFORMITY OF THE SEED SCALE

### A. Smearing budget

The width of the break is itself a measurement. It bounds how uniform the seed scale is across the sources. Because the diffuse break is the superposition of many sources, the observed feature is the single-source threshold shape convolved with the spread of break energies across the population. Writing the smearing of each contribution in $\log_{10} E$ and treating the contributions as independent, the total break width adds in quadrature,

$$\sigma_s^2 = \sigma_{\mathrm{thr}}^2 + \sigma_z^2 + \sigma_\varepsilon^2, \tag{3}$$

with $\sigma_{\mathrm{thr}}$ the intrinsic single-source threshold width, $\sigma_z$ the redshift spread of the population, and $\sigma_\varepsilon$ the dispersion of the seed scale $\varepsilon_0$ across sources [since $E_{\mathrm{br}} \propto [(1+\bar{z})\varepsilon_0]^{-1}$, a spread in either $1+z$ or $\varepsilon_0$ displaces the break by the same logarithmic amount].

### B. Redshift kernel

The redshift term is calculable. Each source places its break at $E_{\mathrm{br}} \propto (1+z)^{-1}$, so $\sigma_z$ is the standard deviation of $\log_{10}(1+z)$ weighted by the diffuse-flux contribution per shell, $w(z) \propto R(z)\,(1+z)^{-\gamma}/H(z)$, with $R(z)$ the comoving source-rate density, $H(z)$ the expansion rate, and $\gamma \simeq 2$ the local index at the peak. For a flat $\Lambda$CDM cosmology [22] ($\Omega_m = 0.3$) and source evolutions spanning the cosmic star-formation history [23] and a stronger active galactic nucleus (AGN)-like evolution peaking near $z \simeq 2$ [24], we find $\sigma_z \simeq 0.15$–$0.19$ dex, with a representative value

$$\sigma_z \simeq 0.16 \text{ dex}, \tag{4}$$

insensitive to the high-redshift cutoff above $z \simeq 3$.

### C. Single-source threshold width

The single-source term $\sigma_{\mathrm{thr}}$ is calculable, and the Appendix computes it. The single-source spectrum is evaluated with the full photopion cross section, the isotropic-target angle average, and the decay kinematics of the secondaries, and $\sigma_{\mathrm{thr}}$ is defined operationally as the Gaussian width for which the smeared broken power law best reproduces the computed curve over the measured band, the same convolution used in the fit below. This fit-based definition matters because the decay kernel is asymmetric, so a naive quadrature of component widths misleads.

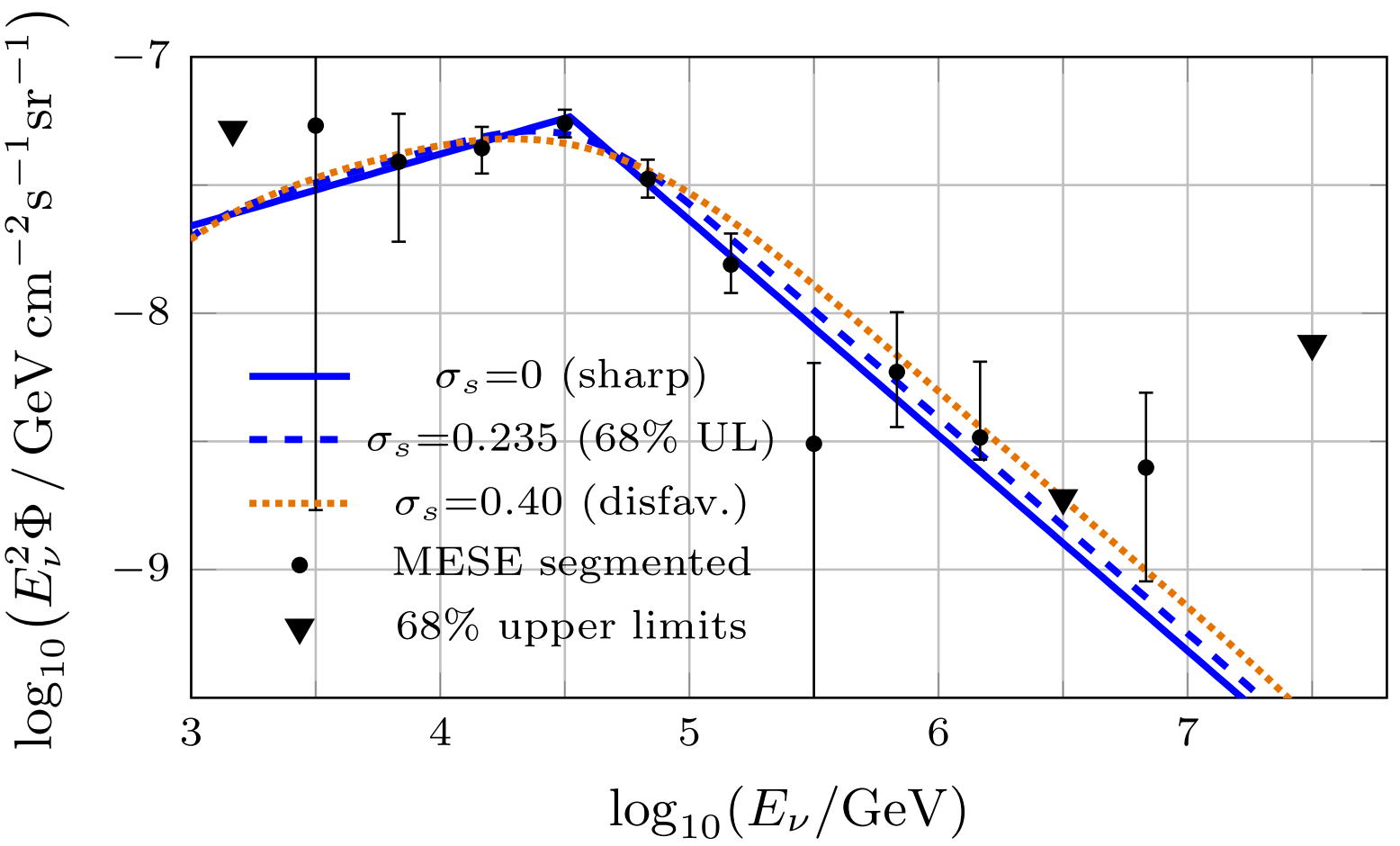

FIG. 4. The broken-power-law fit behind the smearing bound, shown against the official MESE segmented flux of the companion measurement [21]. Points are the per-bin normalizations with $1\sigma$ errors. Downward triangles are 68% upper limits. The solid curve is the sharp BPL ($\sigma_s = 0$) and the dashed curve carries the 68% upper limit on the total smearing ($\sigma_s = 0.235$ dex). The two nearly coincide, since the 0.33-dex segment width leaves smearing below the bin scale unresolved, so the bound is a consistency upper limit rather than a detection. The dotted curve ($\sigma_s = 0.40$ dex) rounds the peak below the break bin and is disfavored.

The result is dominated not by the cross section, whose resonance width and angle average are absorbed into the break position, but by the spread of neutrino energies in the pion and muon decays. For the thermal target, the intrinsically steeper-than-power-law threshold partly compensates the kinematic broadening, and $\sigma_{\mathrm{thr}} \simeq 0.14$ dex (range 0.08–0.20 across the modeling variations of the Appendix), comparable to $\sigma_z$. A sharply truncated power-law target instead gives $\sigma_{\mathrm{thr}} \simeq 0.3$ dex in all variants.

Since $\sigma_{\mathrm{thr}}$ can only add to the kernel, using $\sigma_z$ alone gives the most conservative (weakest) bound below. Subtracting the calculated central $\sigma_{\mathrm{thr}} = 0.14$ dex as well would tighten Eq. (5) to $\sigma_\varepsilon \lesssim 0.11$ dex (factor $\lesssim$1.3), while at the upper end of the modeling range the threshold and redshift kernels alone saturate the measured bound, leaving no room for seed-scale dispersion at all. The conservative bound of Eq. (5) remains the quoted result.

### D. Bound on the seed-scale dispersion

The measured break is sharp. Fitting the smeared broken power law to the official MESE segmented flux [21], with only the normalization and $\sigma_s$ free, bounds the total smearing to $\sigma_s \leqslant 0.235$ dex (68% CL, Fig. 4), barely above the redshift kernel alone. Equation (3) then inverts to

$$\sigma_\varepsilon \leqslant \sqrt{\sigma_s^2 - \sigma_z^2} \simeq 0.17 \text{ dex} \quad (\text{factor } \lesssim 1.5), \tag{5}$$

consistent with $\sigma_\varepsilon = 0$. The seed scale, i.e., the soft-x-ray temperature that sets the photomeson threshold, is therefore uniform across the dominant source population to within a factor $\lesssim$1.5. The sources behave as standard candles in $\varepsilon_0$.

Two caveats temper this. The segments are 0.33 dex wide, so smearing below the bin scale is unresolved. The best-fit total smearing ($\sigma_s \simeq 0.1$ dex) lies below the redshift kernel and is not separately measurable, so Eq. (5) is a consistency upper limit rather than a detection of intrinsic dispersion. It is nonetheless a population-level statement that fits to the mean spectrum do not provide. The observed break is no broader than the irreducible redshift kernel (Fig. 5).

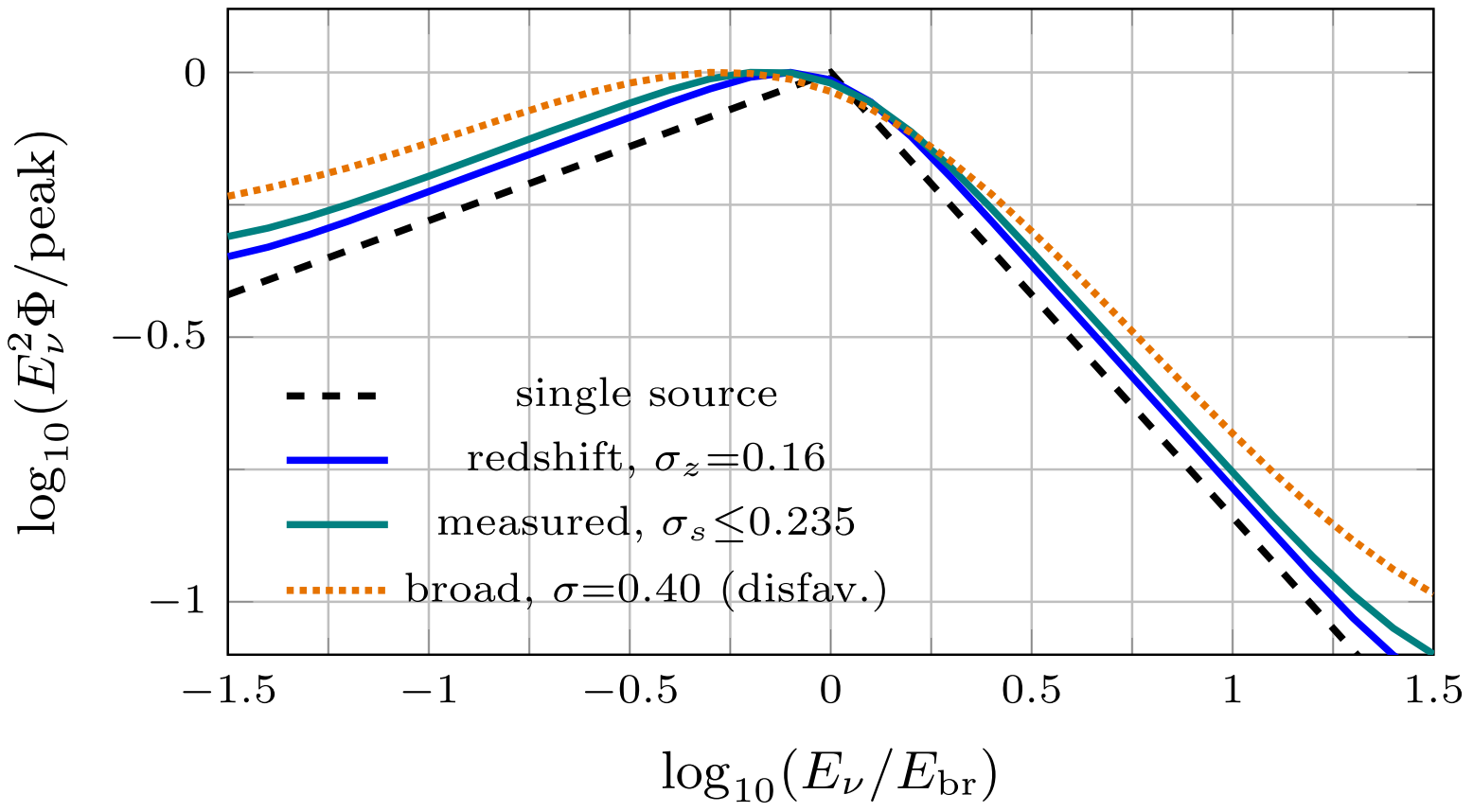

FIG. 5. How sharp the diffuse break stays under population smearing, for the MESE indices ($\gamma_1 = 1.72$, $\gamma_2 = 2.84$). A perfectly sharp single-source threshold (dashed) is convolved in $\log_{10} E_\nu$ with Gaussian kernels of increasing width. The irreducible redshift kernel ($\sigma_z \simeq 0.16$ dex, blue) already rounds the peak. The measured sharpness ($\sigma_s \leqslant 0.235$ dex, teal) lies just above it, whereas a population with a wide spread of seed scales (dotted, 0.40 dex) is disfavored. The narrow margin between the blue and teal curves bounds the seed-scale dispersion [Eq. (5)].

### E. Implications for the source population

The sharpness strengthens the single-source-class reading of the TeV-PeV flux [4]. One homogeneous population with a common seed scale suffices, and a mixture of source classes with different target fields would broaden the break beyond the bound. The full calculation of the Appendix sharpens this. With the decay kinematics included, the thermal target predicts a single-source width ($\simeq$0.14 dex) that fits inside the measured bound alongside the redshift kernel, whereas a sharply truncated power-law target predicts $\simeq$0.3 dex, more rounding than $\sigma_s \leqslant 0.235$ dex permits once $\sigma_z$ is included. The measured sharpness therefore selects the thermal-like cutoff of the warm corona, with the analytic broken power law of Sec. III standing as the effective description of the observable spectrum, whose wings the decay kernel preserves.

It does not, by itself, extend that conclusion to the ultrahigh-energy events. There the evidence points the other way. The resolved sources span over two decades in target scale (Sec. VIII), and KM3-230213A, if it marks a population break, implies a warm-dust target far below the coronal seed (Sec. VII), so the ultrahigh-energy flux traces a population distinct from the one that makes the break.

## V. PREDICTED SOURCE-REDSHIFT DISTRIBUTION

### A. Construction

The forward model that fixes the spectrum also says which source redshifts contribute at each energy, and the typical redshift falls across the break. The integrand of $\Phi(E_\text{obs}) \propto \int dz\,[R(z)/H(z)]\,Q(E_\text{obs}(1+z))$ is the joint density in $(E_\text{obs}, z)$, so for neutrinos observed in an energy band the redshift distribution is $dN/dz \propto [R(z)/H(z)]\int dE_\text{obs}\,Q(E_\text{obs}(1+z))$. This is a population statement, not a per-event map. A naive inversion $E_\text{obs} = \mathcal{C}/(1+z)$ fails because each source emits a broad spectrum rather than a line, and because the seed scale $\varepsilon_0$ is degenerate with $1+z$ in $E_\text{br}$.

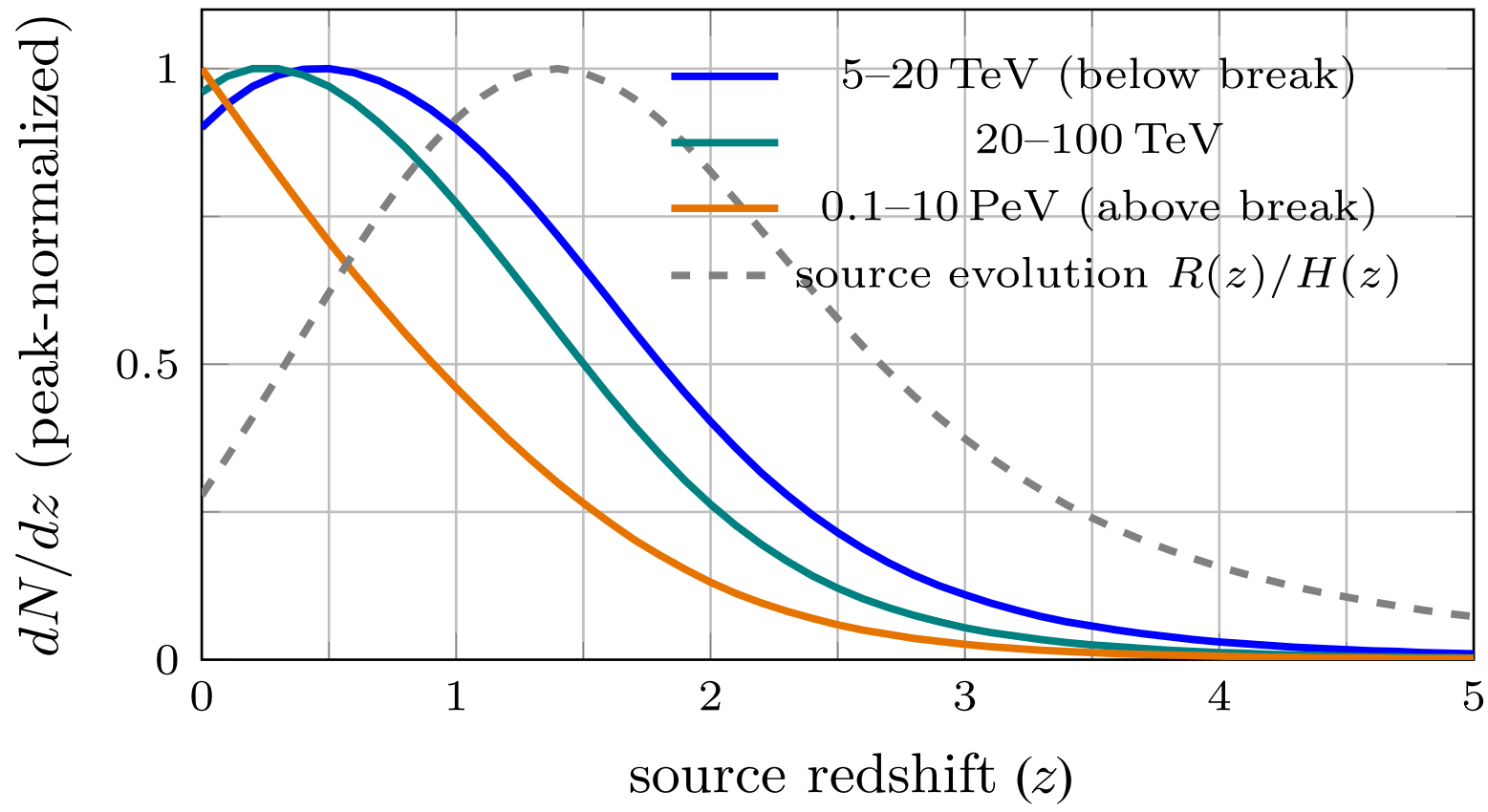

FIG. 6. Predicted source-redshift distribution of the detected neutrinos, shifting to lower redshift above the break, in three observed-energy bands (MESE indices, star-formation evolution, flat ΛCDM, peak normalized). All three colored curves describe one and the same source population. They differ only in the observed neutrino energy selected, not in source class. The gray dashed curve is that population's redshift distribution before any energy selection, $dN/dz \propto R(z)/H(z)$ for the star-formation evolution. The colored curves shift around it because the spectrum weights redshifts differently on the two sides of the break. The typical redshift decreases with observed energy. Neutrinos below the break originate at systematically higher $z$ than those above it, because the steep above-break index suppresses the high-$z$ contributions. The curves are a prediction of the threshold model folded with the source evolution, not an inversion of individual energies to redshifts.

### B. Results and implications

Figure 6 shows $dN/dz$ for three observed-energy bands using the star-formation evolution. Because the steep index $\gamma_2$ above the break suppresses the redshift-stretched, high-$z$ contributions, the typical source redshift falls across the break. For the star-formation evolution $\langle z\rangle \simeq 1.20$ below it (5–10 TeV), $\simeq$1.00 at the break, and $\simeq$0.86 above it ($\gtrsim$100 TeV), where it plateaus once all contributing source energies lie above the break. In the same below/break/above order, the strong AGN-like evolution gives $\langle z\rangle \simeq 2.08/1.77/1.67$ and the nonevolving case $\simeq$0.85/0.64/0.50. The spectral break thus coincides with a transition in contributing redshift. The direction of the shift is robust. Its size, the ratio of the below-break to above-break typical redshift, runs from $\simeq$1.2 (AGN) through $\simeq$1.4 (star formation) to $\simeq$1.7 (nonevolving). The absolute shift is nearly the same in each case, so it is a larger fractional change where the typical redshift is lower. The absolute redshift scale itself tracks the assumed evolution and inherits the $\varepsilon_0$–$z$ degeneracy. What the measured spectrum fixes, given a universal seed scale, is the shape of $dN/dz$ and its energy dependence, not the overall normalization in $z$.

This $dN/dz$ is also an input to diffuse-flux probes of long-baseline new physics. Constraints on quasi-Dirac mass splittings [25] and on neutrino decay [26], both of which imprint energy-dependent effects accumulated over cosmological distances, integrate over a source-redshift distribution that is normally taken to follow the star-formation history. The threshold model instead predicts the shape and energy dependence of that distribution from the measured break, so such bounds can be tested against a spectrum-derived $dN/dz$ rather than an assumed one.

## VI. FLAVOR AND ANTINEUTRINO STRUCTURE

### A. Composition at the source

The same threshold that shapes the spectrum also fixes the flavor content, and the antineutrino channel at the Glashow resonance makes that content a discriminator of the high-energy origin. The full six-species decomposition for this $p\gamma$ break, with its Glashow-resonance discriminator and IceCube-Gen2 projections, was computed by Winter *et al.* [4] on the Glashow-resonance framework of Ref. [6]. We recover its main features analytically from the threshold and add the Earth-propagation treatment below.

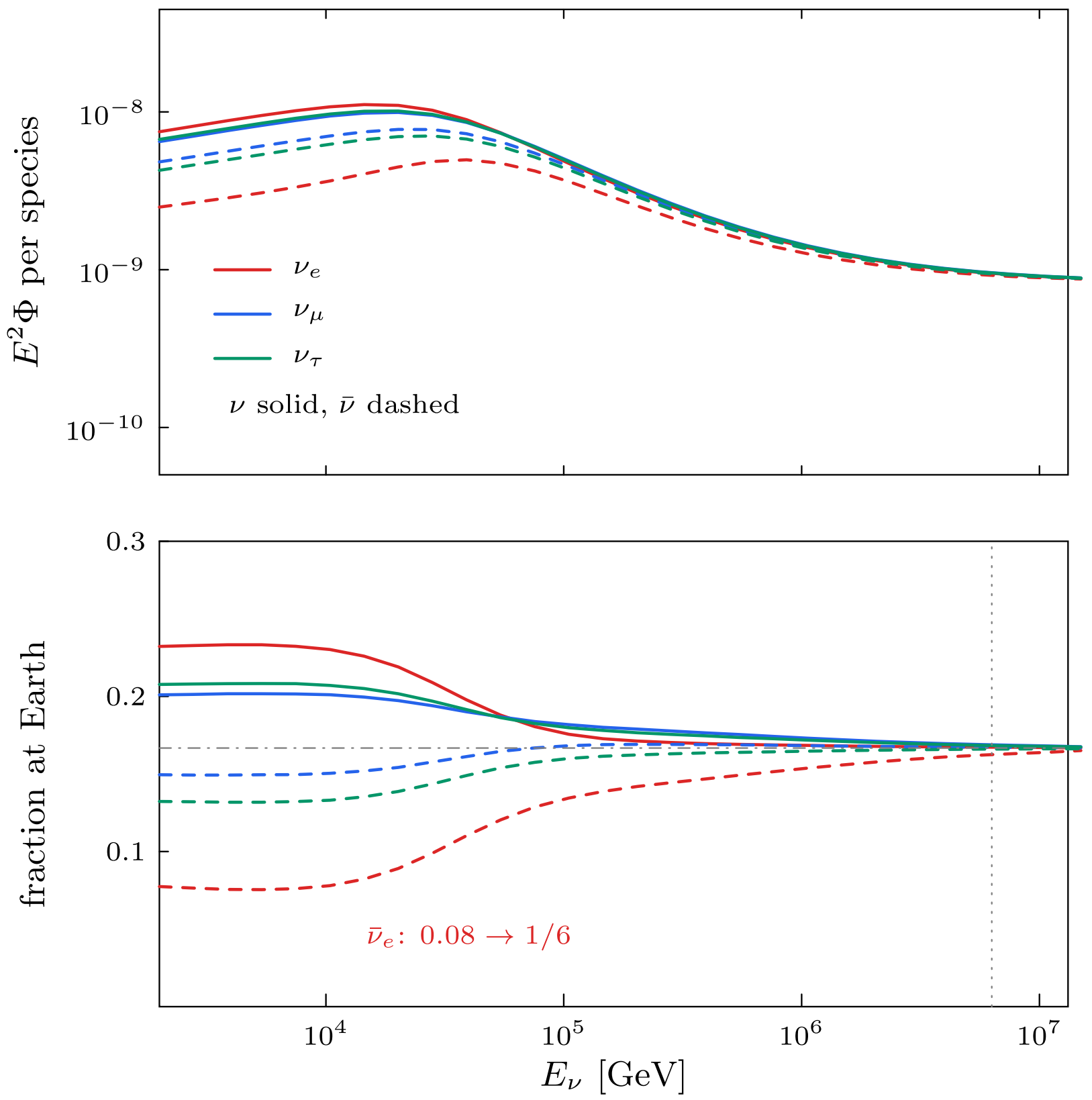

FIG. 7. Six-species decomposition of the incident flux, tracing the $\bar{\nu}_e$ fraction to its Glashow value (MESE parameters, before Earth absorption). Shown is the corona plus a flat, flavor-democratic transparent ($pp$-like) component present across the band. The top panel shows $E^2\Phi$ per species (solid $\nu$, dashed $\bar{\nu}$). The corona falls as $E^{2-\gamma_2}$ onto the flat transparent floor, subdominant near the break and dominant above $\sim$0.5 PeV. The bottom panel shows fractions at Earth. The dash-dotted line is the $pp$ value $1/6$. The corona alone makes $\bar{\nu}_e$ rise from 0.06 to 0.13 as multipion production restores $\pi^-$. With the transparent floor included, the total $\bar{\nu}_e$ runs from $\simeq$0.08 to $1/6$, reaching the $pp$ value and an order-unity rate at the Glashow resonance (6.3 PeV, dotted).

The composition is energy dependent. Below the break, interactions are $\Delta$ dominated and yield $\pi^+$ only. Above it, the multipion plateau produces $\pi^-$ with asymptotic charged-pion fraction $f_{\pi^-} \simeq 0.4$ (charge conservation retains a $\pi^+$ excess). We model $f_{\pi^-} = 0.4\,x^2/(1+x^2)$, $x = E/E_{\rm br}$. Coronal fields $B \simeq 10$ G leave muons undamped throughout the band (synchrotron damping would set in only near $3\times10^3$ PeV, and inverse-Compton damping is suppressed because scattering on the $\sim$100 eV target is deep in the Klein-Nishina regime), so the full chains $\pi^\pm \to \mu^\pm \to e^\pm$ apply.

Propagating through averaged oscillations, the corona component alone has its $\bar{\nu}_e$ fraction rise from 0.06 at 5 TeV through 0.10 at the break to 0.13 in the PeV band, against the flavor-democratic $pp$ value of $1/6$, with all other species within $\simeq$20% of democracy. Including the flat, flavor-democratic transparent ($pp$-like) component, present across the band but subdominant near the break, raises the total $\bar{\nu}_e$ to $\simeq$0.08 at low energy and, as it overtakes the steeply falling corona above $\sim$0.5 PeV, to $1/6$ at the Glashow resonance (Fig. 7). The high-energy flux of interest to the resonance carries the $pp$ composition and an order-unity rate.

The $\nu/\bar{\nu}$ asymmetry is thus concentrated in the electron sector. Standard flavor-ratio observables see a near-canonical pion beam. The corona suppresses its own $\bar{\nu}_e$ at 6.3 PeV mainly through its steep spectrum, which leaves little flux at that energy, and only mildly through its composition (a factor $\simeq$1.2–1.4). The order-unity resonance rate comes instead from the transparent component, which dominates the channel there.

### B. Propagation through the Earth

Earth propagation reshapes what is detectable. With the PREM profile and Cooper-Sarkar-Mertsch-Sarkar cross sections [27], near-horizon trajectories ($\cos\theta_z = -0.1$, column $3.9\times10^8\,\mathrm{g\,cm^{-2}}$) transmit 0.94 at 100 TeV and 0.82 at 1 PeV, but the resonant $\bar{\nu}_e$ cross section ($\simeq 5\times10^{-31}\,\mathrm{cm^2}$ per electron) makes even this chord opaque at the $10^{-26}$ level. Glashow events are detectable exclusively from above the horizon. Steeply upgoing paths ($\cos\theta_z = -0.8$) transmit 0.47 at 100 TeV and 0.08 at 1 PeV. MSW matter effects, computed by full three-flavor evolution of the incoherent mass-eigenstate flux through PREM, are confined to the GeV resonances [$\mathcal{O}(1)$ conversion at 3–10 GeV, in the $\nu$ channel for normal ordering] and decay to max $|P_{\mathrm{Earth}} - P_{\mathrm{vac}}| \simeq 5\times10^{-4}$ at 5 TeV and $3\times10^{-6}$ at 1 PeV [7–9]. The Earth is flavor-transparent throughout the band, and the mass ordering is invisible here.

### C. Confrontation with flavor measurements

These predictions confront IceCube's flavor measurements directly. Because the threshold only exchanges $\nu_e \leftrightarrow \bar{\nu}_e$ within the electron sector, the flavor-summed composition at Earth is *energy independent*, $(f_e : f_\mu : f_\tau) = (0.31 : 0.35 : 0.34)$, identical to the canonical pion beam, so all $\nu/\bar{\nu}$-blind measurements should match standard expectations. They do (Fig. 8). The preliminary high-energy starting events (HESE) 7.5-yr ternary fit gives 0.29 : 0.50 : 0.21 [28], the 12-yr update $0.19^{+0.26}_{-0.15} : 0.43^{+0.27}_{-0.17} : 0.38^{+0.37}_{-0.24}$ with the model point well inside the 68% region (and a spectral index $2.84^{+0.19}_{-0.18}$ coinciding with $\gamma_2$) [29], while the $\nu_\tau$ detections (two double cascades disfavoring zero astrophysical $\nu_\tau$ at $2.8\sigma$ [28], strengthened to seven candidates in 9.7 yr [30]) are consistent with $f_\tau = 0.34$. This agreement is necessary but not discriminating. The discriminating channel is $\bar{\nu}_e$.

### D. The Glashow-resonance rate

The transparent ($pp$-like) component that carries the high-energy flux sets the resonance rate. An exact calculation gives the order-unity $pp$/single-power-law expectation, 0.55–1.1 expected hadronic-channel events in the 4.6-yr search (containment factor 1–2), consistent with the single observed Glashow candidate [17]. The inputs are the peak cross section $\sigma_{\mathrm{pk}} = 4.87\times10^{-31}\,\mathrm{cm^2}$, the Breit-Wigner full width at half maximum of 0.33 PeV, the electron content of a cubic kilometer of ice ($n_e V$), and a $2\pi$ downgoing sky, since the Earth is opaque at the resonance from below. A corona terminating at the break falls $\simeq 10\times$ short, $E^2\Phi_{\bar{\nu}_e}(6.3\,\mathrm{PeV}) \simeq 9\times10^{-11}\,\mathrm{GeV\,cm^{-2}\,s^{-1}\,sr^{-1}}$ and 0.055–0.11 events ($P_{\geqslant 1} \simeq 5\%$–$10\%$), and a corona truncated at $E_{p,\mathrm{max}}$ supplies none. The candidate's presence therefore already favors a transparent high-energy component.

A decade of IceCube-Gen2 ($\sim 8\,\mathrm{km^3}$) sharpens the test, giving $\simeq$9–14 hadronic Glashow events for the transparent ($pp$-like) prediction, $\simeq$1 for a corona without cutoff, and $\simeq$0 for the cutoff variant, distinctions ternary measurements cannot make. Figure 9 shows the two-zone composition. The corona supplies the spectral bump in every flavor, while the transparent component, flavor democratic and hard, dominates the $\bar{\nu}_e$ channel above $\sim$0.5 PeV and carries the Glashow flux the corona lacks.

## VII. KM3-230213A AND THE ULTRAHIGH-ENERGY LANDSCAPE

The relation does not stop at the break. The KM3NeT/ARCA (Cubic Kilometre Neutrino Telescope/Astroparticle Research with Cosmics in the Abyss) event KM3-230213A, with median energy 220 PeV (90% CL, 72 PeV–2.6 EeV) [31], lies far above it and tests how far the photomeson picture extends. The coronal component extrapolates to $E^2\Phi \simeq 3.5\times10^{-11}$ GeV cm$^{-2}$ s$^{-1}$ sr$^{-1}$ at 220 PeV (zero if the few-hundred-TeV deficit marks the corona $E_{p,\max}$), so the model is consistent with IceCube's 12.6-yr extremely-high-energy (EHE) null and the absence of a directional excess at the location [31,32]. The 1.6–2.9$\sigma$ joint tension [33] concerns a distinct component, favored there as two-component over a single power law and bounded by the joint normalization $E^2\Phi^{1f} = 7.5\times10^{-10}$ GeV cm$^{-2}$ s$^{-1}$ sr$^{-1}$.

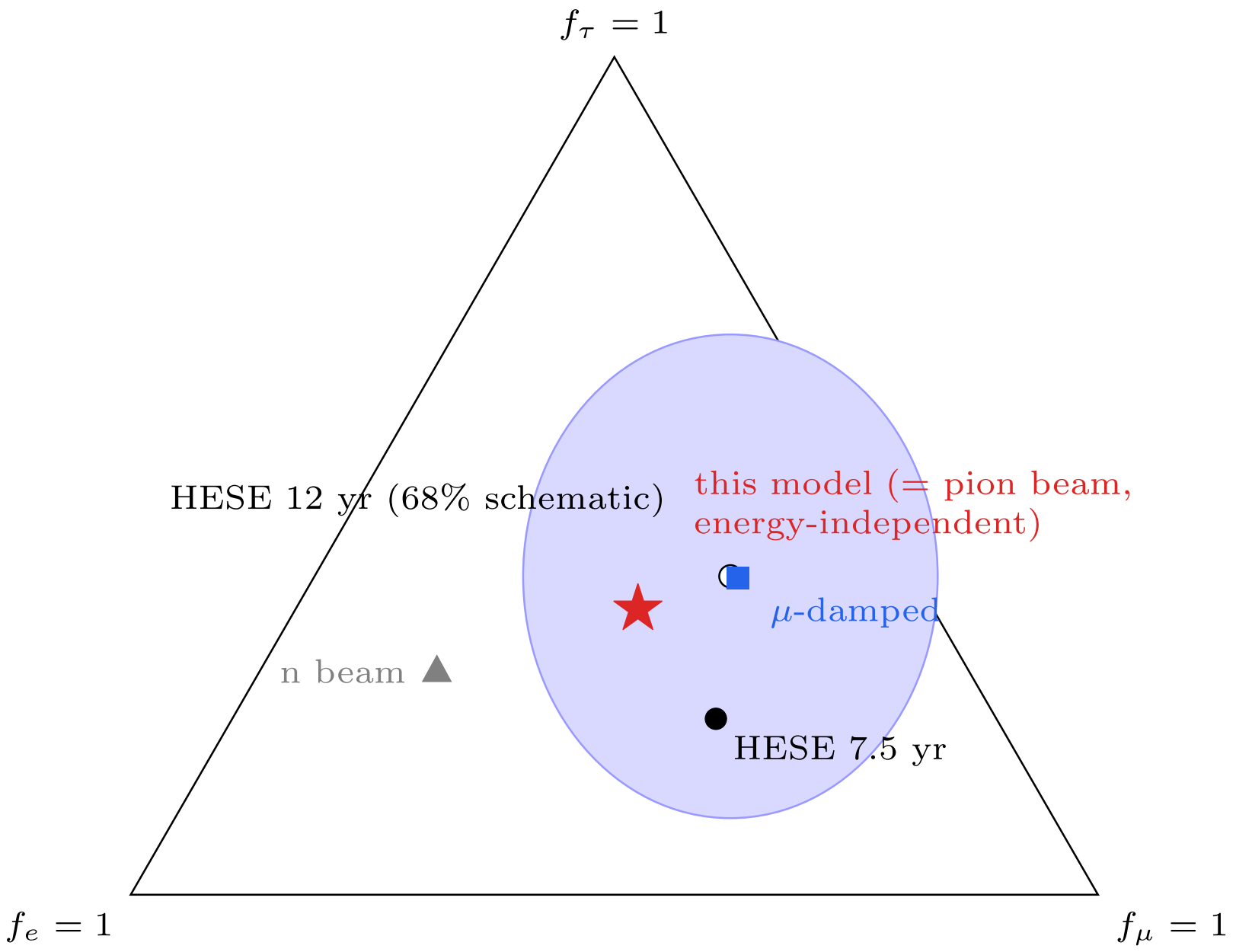

FIG. 8. Flavor composition at Earth, with the model inside the IceCube allowed region. The model point (star) is energy independent at (0.31 : 0.35 : 0.34), coinciding with the canonical pion beam, because the threshold only exchanges $\nu_e \leftrightarrow \bar{\nu}_e$. The $\pi^+$ and $\pi^-$ chains share the 1 : 2 : 0 flavor ratio, so the rise of $\pi^-$ above the break shifts strength between $\nu_e$ and $\bar{\nu}_e$ without altering $e : \mu : \tau$. IceCube HESE best fits [28,29] are shown with a schematic 68% region constructed from the quoted one-dimensional intervals of the 12-yr fit (the published contour is not shown here). Muon-damped and neutron-beam benchmarks are shown for comparison. All $\nu/\bar{\nu}$-blind measurements are consistent with the model by construction. The discriminating information resides in the $\bar{\nu}_e$ sector.

If the event marks a population break, Eq. (2) inverts to $\varepsilon_0 \simeq 0.013$–$0.04$ eV, warm dust ($T \simeq 150$–$460$ K) in infrared-luminous environments. A cosmogenic reading [34] within the upper energy interval is pressured by the neutrino-based disfavoring of a pure-proton ultrahigh-energy cosmic ray (UHECR) composition [32].

Extended to $E_{p,\max} \sim$ few EeV, the transparent component reaches this bound at 220 PeV (Fig. 3), where IceCube expects $\lesssim$1 EHE event, leaving KM3-230213A a $\sim$2$\sigma$ fluctuation. The two events are not on equal footing, since making the KM3 event typical would give $\sim$70–90 EHE events [32,33] against zero ($>10\sigma$), while the Glashow rate sits at the measured flux. The shared normalization is not a shared mechanism. The Glashow channel reads $\bar{\nu}_e$ resonantly at 6.3 PeV from above the horizon. KM3-230213A is instead a deep-inelastic $\nu_\mu$ track at 220 PeV, some 35 times above the resonance, where $\bar{\nu}_e$ is just one flavor among six species and would appear as a shower rather than a track. Its slightly upgoing arrival is also the wrong hemisphere for the resonance.

If KM3-230213A is instead transient, the lowest-tension reading at $\sim$2$\sigma$, the candidates in its field are flaring blazars [35] and a super-Eddington accretion flare of the nearest blazar

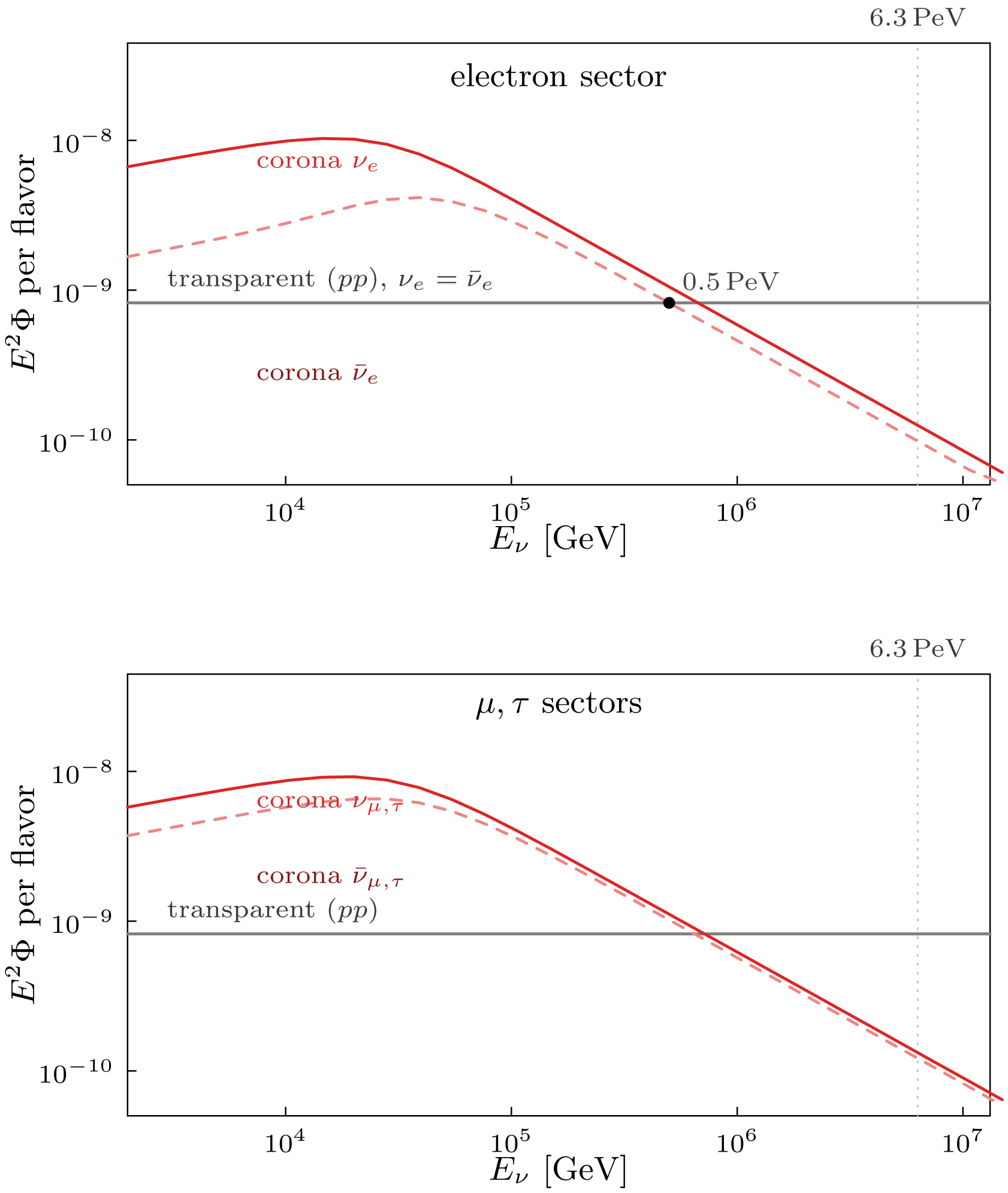

FIG. 9. Two-zone decomposition of the per-flavor diffuse flux at Earth, separating the corona bump from the transparent component that carries the Glashow flux. In each panel, $\nu$ is a solid line and $\bar{\nu}$ a lighter dashed line. The top panel shows the electron sector. The soft corona component (red) is the photomeson-threshold prediction of Fig. 3. It is $\pi^+$ dominated, so its $\bar{\nu}_e$ holds the suppressed 0.06–0.13 fraction and falls steeply above the break. The bottom panel shows the $\mu$ and $\tau$ sectors, clustered near democracy. At the source, each $\pi^\pm \to \mu^\pm$ chain emits $\nu_\mu$ and $\bar{\nu}_\mu$ in equal numbers and $\nu_\tau = \bar{\nu}_\tau = 0$. Averaged oscillations, approximately $\mu$-$\tau$ symmetric and identical for $\nu$ and $\bar{\nu}$, hold the four curves to a few percent in $\mu$-$\tau$, but mix the $\pi^+$-excess electron asymmetry into these sectors through $P_{\mu e}$ and $P_{\tau e}$, leaving a residual $\nu/\bar{\nu}$ split of $\simeq 10\%$ ($\nu_{\mu,\tau} \gtrsim \bar{\nu}_{\mu,\tau}$). The asymmetry is largest in the electron sector. The hard transparent component (gray) is the $pp$-like outflow, flavor democratic ($\bar{\nu}_e = \nu_e = 1/6$ from equal $\pi^+/\pi^-$ and undamped muons) and hard ($E^2\Phi$ flat) to past 6.3 PeV. Its normalization is fixed to the high-energy flux, an order-unity Glashow rate, and is not predicted by the corona model. Above $\sim$0.5 PeV, the transparent component dominates the $\bar{\nu}_e$ channel ($\simeq 10\times$ at 6.3 PeV) and supplies the Glashow candidate, which the corona alone cannot.

MRC 0614–083, whose photomeson target is a delayed infrared echo in the surrounding dust torus [36]. That realizes the warm-dust target on a jetted, flaring AGN rather than an x-ray Seyfert corona, matching our expectation that the ultrahigh-energy events trace a different population from the break.

TABLE II. Per-source test of the break relation, showing the resolved sources span over two decades in target scale. Equation (6) converts each secure IceCube identification's redshift and characteristic neutrino energy into an implied target photon scale $\varepsilon_0$. The steady Seyfert-like sources imply keV x-ray targets and the high-energy TXS 0506+056 alert an ultraviolet one, so the resolved sources are not a single seed-scale population.

| Source ($z$) | $E_\nu^{\rm obs}$ | $\varepsilon_0$ | Target |
|---|---|---|---|
| NGC 1068 (0.0038) | 1.5–15 TeV | $\gtrsim$keV | X-ray corona |
| PKS 1424+240 (0.605) | ~few TeV | $\gtrsim$keV | X-ray corona |
| TXS 0506+056 (0.337) | 290 TeV | $\simeq$20 eV | Jet UV |
| 2014–15 flare | 30–100 TeV | 0.06–0.2 keV | Soft x-ray |

## VIII. RESOLVED NEUTRINO SOURCES

### A. Per-source seed energies

Read as a constraint on a single source, the break relation turns each multimessenger identification into a measurement of its target photon scale. Inverting Eq. (2) for a source at redshift $z$ with characteristic observed neutrino energy $E_\nu^{\rm obs}$ defines the *seed energy* implied by a detection,

$$\boxed{\varepsilon_0 \simeq 0.13\,\mathrm{keV}\left(\frac{E_\nu^{\rm obs}}{33\,\mathrm{TeV}}\right)^{-1}\left(\frac{1+z}{2}\right)^{-1} \simeq \frac{8.6\,\mathrm{keV}}{(E_\nu^{\rm obs}/\mathrm{TeV})\,(1+z)}.} \tag{6}$$

Table II applies this seed-energy relation to the three secure IceCube point sources [20,37,38].

Every entry maps to a physically reasonable target. NGC 1068 and PKS 1424+240 have soft TeV spectra with no resolved turnover, so their break lies at or below the lowest detected energy and Eq. (6) returns $\varepsilon_0 \gtrsim$ keV, the hard x-ray corona expected for a Seyfert. The high-energy TXS $0506+056$ alert returns $\varepsilon_0 \simeq 20$ eV, the ultraviolet external field of a jetted blazar, while its 2014–15 flare [39] sits between soft x-ray and extreme ultraviolet. None of the identifications requires an unphysical target.

### B. Outliers and the break population

The implied seeds span more than two decades, from ~20 eV to several keV, so the resolved sources are not a standard-candle set. A diffuse break uniform to a factor $\lesssim$1.5 (Sec. IV) cannot be assembled from them. It requires a separate, homogeneous population, fainter and more distant Seyferts near $z \sim 1$, with a $\simeq$0.13 keV soft-excess seed, on which the resolved objects sit as bright outliers. NGC 1068 and PKS 1424+240 confirm the photomeson-on-x-rays mechanism but at a harder seed than the diffuse population, marking them as nearby outliers rather than templates. The soft, coronalike spectrum of PKS 1424+240, nominally a BL Lac, is consistent with a hidden Seyfert in the same field.

The high-energy identifications point instead to ultraviolet targets, the jetted and transient class to which KM3-230213A also belongs (Sec. VII). Three tidal-disruption events (AT2019dsg, AT2019fdr, AT2019aalc) were tentatively linked to IceCube alerts but fall outside the revised containment regions of IceCat-2 [40], and the blazar GB6 J1542+6129 has a redshift constrained only to $0.34 \leqslant z \leqslant 1.76$, so neither anchors Eq. (6). The defensible set is the three sources of Table II.

Equation (6) thus doubles as a diagnostic seed-energy plane (Fig. 10). Iso-$\varepsilon_0$ contours partition the $(E_\nu^{\rm obs}, 1+z)$ plane into target bands, and a redshift-resolved source or event falls into the band of its production environment. Three limits keep this a diagnostic rather than a per-event classifier. It requires a redshift. Without one, only the product $(1+z)\varepsilon_0$ is fixed, and the anonymous diffuse flux yields a population-averaged seed (Sec. V), not a per-event label. A single event energy is not the break energy. It is one draw from a spectrum hard below the break and soft above, so for a lone high-energy alert Eq. (6) returns a one-sided bound rather than a measurement, and the classification is into broad bands. And it presupposes a photomeson origin. A transparent $pp$ source has no threshold and no $\varepsilon_0$, for which the absence of a break is itself the discriminant, as in consequence (1) below.

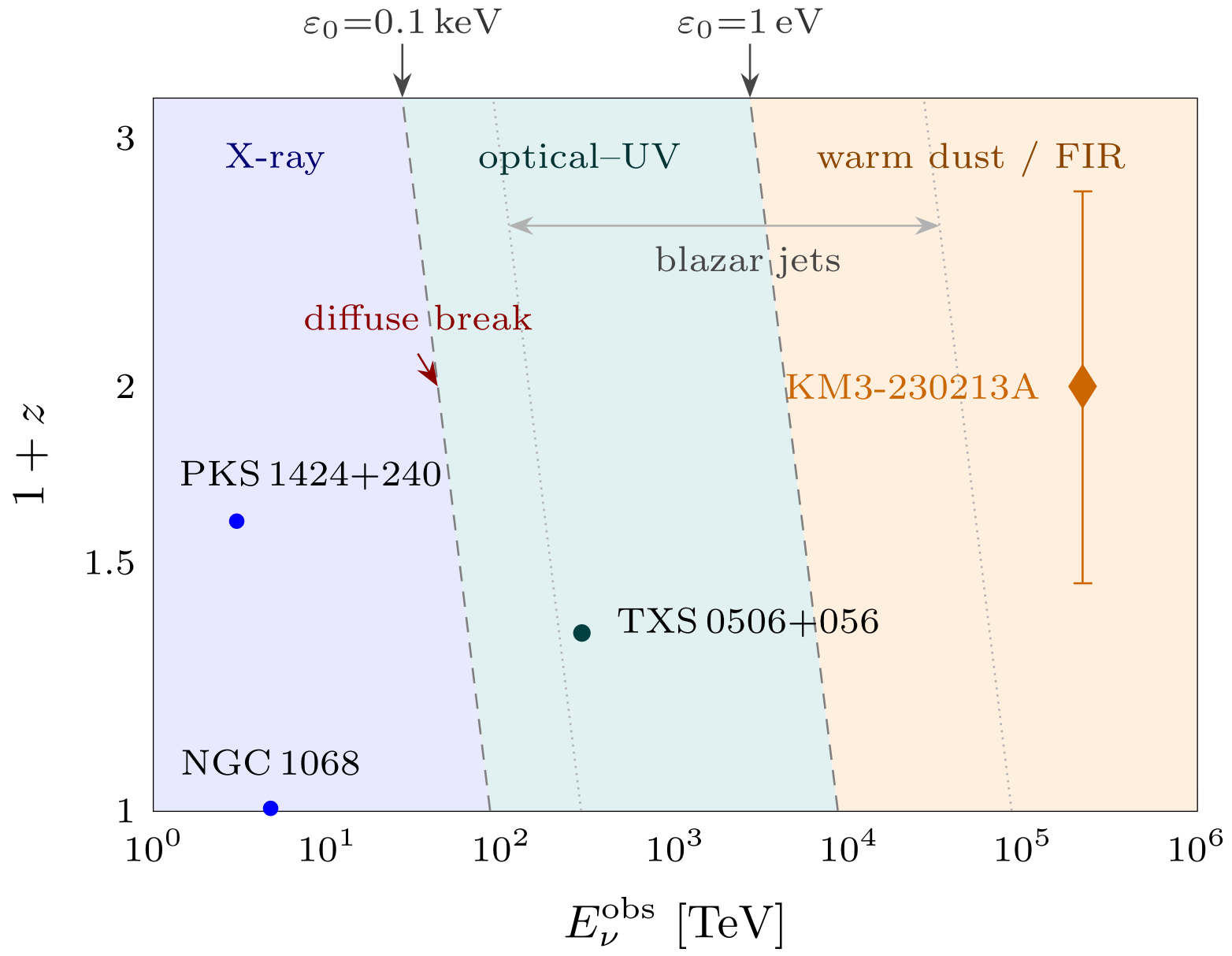

FIG. 10. The seed-energy plane reads each source's target scale from its observed energy and redshift. Iso-$\varepsilon_0$ contours of Eq. (6) (dashed) partition the observed-energy-redshift plane into target-environment bands, x-ray corona ($\varepsilon_0 \gtrsim 0.1$ keV), optical-ultraviolet (1 eV–0.1 keV), and warm dust or FIR ($\varepsilon_0 \lesssim 1$ eV). The secure resolved sources (blue circles), the diffuse break at $\bar{z} \simeq 1$ (arrow), and KM3-230213A (orange diamond at a representative redshift, with a tick spanning the plausible range, since its $z$ is unknown) fall into the bands of their production environments. The dotted band marks where blazar jets land as the jet samples broad-line or disk ultraviolet (∼1–30 eV) down to a dusty-torus infrared echo (∼0.1 eV). Extreme high-frequency-peaked BL Lac synchrotron (keV) instead pushes some blazars into the x-ray band. A source with a redshift maps to a target band. Without one, only the product $(1+z)\varepsilon_0$ is fixed. The plane shows photomeson-threshold tracers only, so an event fixed at the $W$ resonance, such as the 6.3 PeV Glashow $\bar{\nu}_e$, is excluded by construction.

## IX. OBSERVATIONAL CONSEQUENCES

The threshold reading makes six predictions, each testable with current or near-term data.

(1) The hard sub-break spectrum ($E^{-\gamma_1}$) should continue into the 1–5 TeV band without flattening. A $pp$ (proton-gas) component carries no threshold and would soften it to $E^{-s}$ there, so the observed hardness already limits the gas in the proton zone (Thomson depth $\lesssim 0.2$). A flattening to $E^{-s}$ at 1–3 TeV would instead measure the gas density of the acceleration region.

(2) The Glashow rate at 6.3 PeV is the order-unity $pp$ value, set by the transparent component that carries the high-energy flux. A corona truncating at the break would fall several times short, so the $\bar{\nu}_e$ channel discriminates the high-energy origin and is the sharpest compositional test.

(3) If a muon-damped envelope component enters at the band edge (source composition 0 : 1 : 0, Earth ≃1 : 1.8 : 1.8 [41]), the flavor composition drifts across the corona/envelope seam, a Gen2 observable.

(4) Break sharpness should remain consistent with a population-smeared sharp feature, and per-source breaks should scatter as $(1+z)^{-1}\varepsilon_0^{-1}$.

(5) Ultrahigh-energy events should correlate with far-infrared (FIR) or UHECR-associated catalogs, not with the x-ray-selected Seyferts that dominate below the corona break. The factor $\lesssim$1.5 uniformity of the seed scale (Sec. IV) further disfavors a blazar contribution to the break itself, since the heterogeneous target fields of the blazar sequence would smear it, as the resolved sources confirm by spanning over two decades in seed energy (Sec. VIII; Fig. 10). This is consistent with the homogeneous Seyfert-corona identification. A broader electromagnetic program (x-ray, submillimeter, radio, MeV-TeV $\gamma$-ray, and cosmic-ray diagnostics) can further test the source identification.

(6) The source-redshift distribution falls across the break (Fig. 6), below-break neutrinos arising at systematically higher $z$ than those above. This energy-dependent $dN/dz$ is the source-distance input that diffuse-flux tests of quasi-Dirac oscillations and neutrino decay otherwise assume (Sec. V), so those bounds can be tied to the measured spectrum. The shift toward lower $z$ at high energy does not ease per-source identification. The steep above-break spectrum leaves few events there, the near-isotropic flux indicates an abundant faint population, and the corona class is opaque to its own GeV-TeV $\gamma$ rays, so population-level correlation, as in (5), remains the practical route.

## X. CONCLUSIONS

One kinematic threshold accounts for the break near 30 TeV, the $\Delta(1232)$ resonance of accelerated protons on the soft-x-ray field of accreting black holes, and the same relation generates every result that follows. Closed-form threshold relations invert the four broken-power-law fit parameters to the source physics with no remaining freedom, fixing the seed scale at $\varepsilon_0 \simeq 0.13$ keV and tying the two spectral indices to the proton and target-photon spectra.

Read as a population feature, the width of the break bounds the seed-scale dispersion to a factor $\lesssim$1.5 once the calculable threshold and redshift kernels are removed, a constraint the mean-spectrum fits do not supply. The same population fixes the energy dependence of the source-redshift distribution that diffuse-flux oscillation and decay tests take as input. The six-species flavor and neutrino-antineutrino structure follows from the same threshold. The flavor-summed composition holds at the canonical pion-beam value, while the $\bar{\nu}_e$ channel at the Glashow resonance separates a transparent high-energy component from a corona that truncates at the break.

Applied source by source, the seed-energy relation places the secure IceCube identifications across more than two decades in target energy, from keV coronae (NGC 1068, PKS 1424+240) to the $\sim$20 eV jet of TXS 0506+056, so the resolved sources are not a standard-candle set and the homogeneous break calls for a separate, fainter Seyfert-corona population. KM3-230213A sits at the warm-dust end. The picture is falsifiable with near-term data, through the sub-break hardness, the IceCube-Gen2 Glashow rate, and the predicted scatter and electromagnetic associations of the resolved sources.

## ACKNOWLEDGMENTS

V.B. gratefully acknowledges support from the U.S. Department of Energy, Office of Science, Office of High Energy Physics, under Award No. DE-SC0017647 and from the William F. Vilas Trust Estate.

## DATA AVAILABILITY

The analysis uses published data products. The broken-power-law parameters and the segmented-flux points are taken from the IceCube measurements of Refs. [2,21], and the energy and arrival direction of KM3-230213A from Ref. [31].

## APPENDIX: COMPUTATIONAL DETAILS

The three quantitative results, the redshift kernel $\sigma_z$, the smearing bound $\sigma_s$, and the source-redshift distribution $dN/dz$, are reproduced as follows.

### 1. Cosmology and source evolution

We adopt a flat ΛCDM cosmology [22], $H_0 = 70\ \mathrm{km\,s^{-1}\,Mpc^{-1}}$, $\Omega_m = 0.3$, and $\Omega_\Lambda = 0.7$, with $H(z) = H_0\sqrt{\Omega_m(1+z)^3 + \Omega_\Lambda}$.

The comoving source-rate density $R(z)$ is taken in three forms. The first is the star-formation history $R_{\mathrm{SFR}} \propto (1+z)^{2.7}/[1 + ((1+z)/2.9)^{5.6}]$ [23]. The second is a stronger AGN-like evolution $R_{\mathrm{AGN}} \propto (1+z)^5$ for $z \leqslant 2$ and $\propto (1+z)^{-1.5}$ for $z > 2$ (continuous at $z = 2$), motivated by the x-ray AGN luminosity-density history [24]. The third is a nonevolving case $R = \mathrm{const}$. Redshift integrals run to $z_{\max} = 8$ and are stable above $z \simeq 3$.

### 2. Redshift kernel

A source at redshift $z$ sets its break at $E_{\mathrm{br}} \propto (1+z)^{-1}$, so its contribution to the diffuse flux at fixed observed energy carries weight $w(z) \propto R(z)\,(1+z)^{-\gamma}/H(z)$, with $\gamma$ the local spectral index ($\gamma \simeq 2$ at the peak). The kernel is the dispersion of $u \equiv \log_{10}(1+z)$ under $w$,

$$\sigma_z^2 = \frac{\int dz\, w\,(u - \langle u\rangle)^2}{\int dz\, w}, \qquad \langle u\rangle = \frac{\int dz\, w\, u}{\int dz\, w}. \tag{A1}$$

This gives $\sigma_z = 0.15$–$0.19$ dex across the three evolutions (0.16 dex for $R_{\mathrm{SFR}}$), shifting by $\lesssim$0.01 dex between $\gamma = 2$ and $\gamma = 2.5$.

### 3. Break smearing and the $\sigma_s$ fit

A population spread in break energy broadens the feature. We convolve the sharp single-source shape, $E^2\Phi \propto (E/E_{\mathrm{br}})^{2-\gamma_1}$ for $E < E_{\mathrm{br}}$ and $(E/E_{\mathrm{br}})^{2-\gamma_2}$ for $E > E_{\mathrm{br}}$, with a Gaussian of width $\sigma_s$ in $\log_{10} E$ applied to the linear flux. To bound $\sigma_s$, we fix $(\gamma_1, \gamma_2, \log_{10} E_{\mathrm{br}}/\mathrm{GeV}) = (1.72, 2.84, 4.52)$ to the published MESE values and float only the normalization and $\sigma_s$.

The data are the 13 segmented-flux bins of Ref. [21]. Each tabulated normalization $\phi_i$ (units $10^{-18}\ \mathrm{GeV^{-1}\,cm^{-2}\,s^{-1}\,sr^{-1}}$, an $E^{-2}$ segment referenced at 100 TeV) gives $E^2\Phi_i = \phi_i \times 10^{-8}\ \mathrm{GeV\,cm^{-2}\,s^{-1}\,sr^{-1}}$. Detection bins enter a $\chi^2$ as Gaussian terms in $\log_{10}(E^2\Phi)$ with the tabulated asymmetric errors symmetrized in log space. The three bins with best-fit normalization zero enter as one-sided upper-limit constraints. The normalization is profiled analytically and the 68% CL follows from $\Delta\chi^2 = 1$.

The result is a best fit $\sigma_s \simeq 0.10$ dex and $\sigma_s \leqslant 0.235$ dex (68% CL), with $\chi^2 \simeq 5.9$ over the ten detection bins. This is a simplified profile fit to the tabulated per-bin central values and $1\sigma$ ranges. It does not use the experiment's full likelihood or the bin-to-bin covariance, and the 0.33-dex segment width sets the floor on resolvable smearing. Two detection bins have flux consistent with zero, hence an unbounded lower error in log space. Weighting them by their upper error gives the quoted bound, whereas a more conservative treatment of the lower error loosens it to $\sigma_s \leqslant 0.27$ dex ($\sigma_\varepsilon \leqslant 0.22$ dex, factor $\lesssim 1.7$).

### 4. Single-source threshold width

The width $\sigma_{\mathrm{thr}}$ entering Eq. (3) is computed from the full single-source spectrum. The photopion cross section is modeled as the $\Delta(1232)$ Breit-Wigner (peak 0.50 mb at $\varepsilon' = 0.34$ GeV, width $\Gamma = 0.117$ GeV, threshold 0.15 GeV) plus a multipion plateau (0.12 mb with a smooth turn-on near 0.6 GeV), at the level of the simplified models of Ref. [15]. The interaction rate in an isotropic target follows from the standard angle-averaged integral $R(\gamma_p) \propto \gamma_p^{-2} \int d\varepsilon\, [n(\varepsilon)/\varepsilon^2] \int^{2\gamma_p\varepsilon} d\varepsilon'\, \varepsilon'\sigma(\varepsilon')$, evaluated for the truncated power-law target of Sec. III ($\alpha_x = 2.12$) and for a blackbody at $k_B T = 0.12$ keV. The computed asymptotic slopes reproduce $2 - \gamma_1$ and $2 - \gamma_2$ exactly.

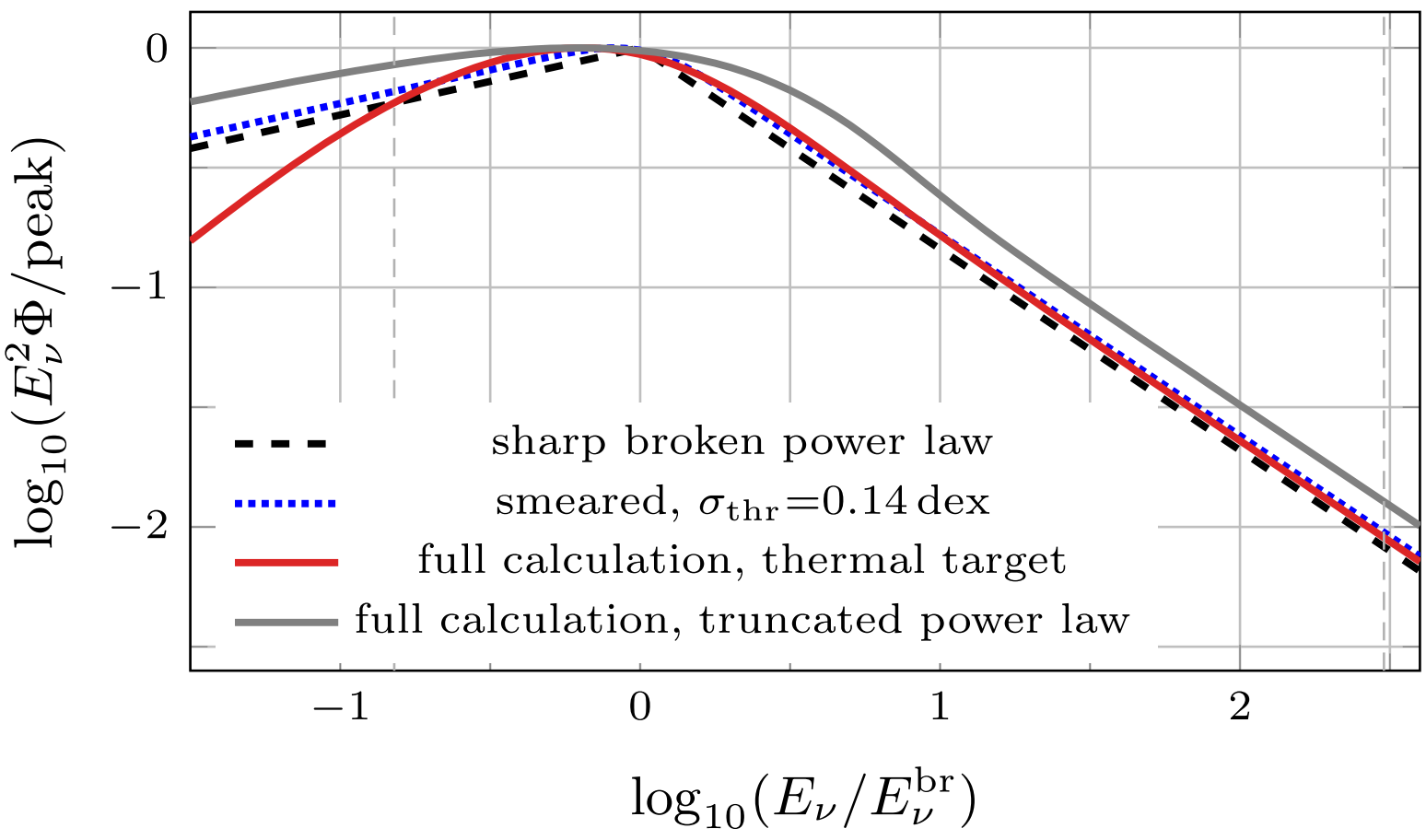

FIG. 11. Single-source threshold width from the full calculation. The computed single-source spectrum for the thermal target (red), with the photopion cross section, the isotropic-target angle average, and the secondary decay kinematics, is compared with the sharp broken power law (dashed) and with the same shape smeared by the best fit $\sigma_{\rm thr} = 0.14$ dex (dotted). The sharp curve already tracks the calculation to 0.03 dex rms over the fitted band (between the vertical marks). The truncated power-law target with the same kinematics (gray) is broader, $\sigma_{\rm thr} \simeq 0.3$ dex, more rounding than the measured $\sigma_s \leqslant 0.235$ dex permits once the redshift kernel is included, so the measured sharpness favors the thermal cutoff.

The neutrino spectrum is $E_p^{-s} R(E_p)$ ($s = 2.84$) convolved with the all-flavor decay kernel of the chain $\Delta^+ \to n\pi^+$ (isotropic decay, $E_\pi/E_p$ uniform on 0.03–0.40), $\pi^+ \to \mu^+\nu_\mu$ ($E_\nu/E_\pi$ uniform on 0–0.43), and $\mu^+ \to e^+\nu_e\bar{\nu}_\mu$ (relativistic unpolarized decay spectra), sampled by Monte Carlo. The kernel has a logarithmic standard deviation of 0.54 dex with a long low-energy tail. Then, $\sigma_{\rm thr}$ is the width for which the smeared broken power law of the fit above best reproduces the computed curve over the source-frame band corresponding to the measurement (normalization and break position free, uniform weights in $\log_{10} E$).

The resonance width and angle average alone give $\sigma_{\rm thr} < 0.03$ dex. The decay kernel dominates. For the thermal target, the result is $\sigma_{\rm thr} \simeq 0.14$ dex (Fig. 11), with the sharp shape itself accurate to 0.03 dex rms. Varying the $\Delta$-decay angular distribution between the isotropic baseline and $(1 + 3\cos^2\theta^*)$-type weightings, the multipion normalization by $\pm 25\%$, the turn-on between 0.5 and 0.8 GeV, and adding a direct-channel proxy spans $\sigma_{\rm thr} \simeq 0.08$–$0.20$ dex.

For the truncated power-law target, whose prekernel shape is already the sharp broken power law, the same kernel gives $\sigma_{\rm thr} \simeq 0.29$–$0.36$ dex in all variants, above the room $\sqrt{\sigma_s^2 - \sigma_z^2} \simeq 0.17$ dex the measurement allows. Three approximations are made. The resonance-region kernel is applied at all energies (above the break the true multiplicities are higher, which mainly renormalizes the soft wing, and the fit floats that normalization). Muon polarization is neglected. The neutron $\beta$ decay $\bar{\nu}_e$, at $E_\nu/E_p \sim 10^{-3}$, is irrelevant near the break.

### 5. Source-redshift distribution

The diffuse flux is

$$\Phi(E_{\rm obs}) \propto \int_0^{z_{\rm max}} dz\, \frac{R(z)}{H(z)}\, Q(E_{\rm obs}(1+z)), \tag{A2}$$

with the source-frame per-source spectrum a smoothly broken power law,

$$Q(E') = E'^{-\gamma_1}[1 + (E'/E'_{\rm br})^{1/\delta}]^{-(\gamma_2-\gamma_1)\delta}, \tag{A3}$$

using $(\gamma_1, \gamma_2) = (1.72, 2.84)$, smoothness $\delta = 0.1$, and a source-frame break $E'_{\rm br}$. Since $E_{\rm br} \propto [(1+\bar{z})\varepsilon_0]^{-1}$, $E'_{\rm br}$ is degenerate with $\varepsilon_0$ and the mean redshift. We set $E'_{\rm br} \simeq 66$ TeV, so the stacked observed break falls at the measured $\simeq$33 TeV for $R_{\rm SFR}$. The redshift distribution of neutrinos detected in a band $[E_a, E_b]$ is

$$\frac{dN}{dz} \propto \frac{R(z)}{H(z)} \int_{E_a}^{E_b} dE_{\rm obs}\, Q(E_{\rm obs}(1+z)), \tag{A4}$$

whose band means give the $\langle z \rangle$ values of Sec. V. The absolute redshift scale tracks the assumed evolution and the $\varepsilon_0$–$z$ degeneracy. The spectrum fixes only the shape of $dN/dz$ and its energy dependence given a universal seed scale.

---

[1] M. G. Aartsen, R. Abbasi, Y. Abdou, M. Ackermann, J. Adams, J. A. Aguilar, M. Ahlers, D. Altmann, J. Auffenberg, X. Bai, *et al.* (IceCube Collaboration), Evidence for high-energy extraterrestrial neutrinos at the IceCube detector, Science **342**, 1242856 (2013).

[2] R. Abbasi, M. Ackermann, J. Adams, S. K. Agarwalla, J. A. Aguilar, M. Ahlers, J. M. Alameddine, S. Ali, N. M. Amin, K. Andeen, *et al.* (IceCube Collaboration), Evidence for a spectral break or curvature in the spectrum of astrophysical neutrinos from 5 TeV to 10 PeV, Phys. Rev. Lett. **136**, 121002 (2026).

[3] A. Khatee Zathul, K. Fang, F. Halzen, and D. Hooper, The Delta resonance in the neutrino sky, arXiv:2606.00249.

[4] W. Winter, D. F. G. Fiorillo, and S. Buson, Single-source-class interpretation of the diffuse astrophysical neutrino flux, Phys. Rev. D **113**, 123043 (2026).

[5] S. L. Glashow, Resonant scattering of antineutrinos, Phys. Rev. **118**, 316 (1960).

[6] V. Barger, L. Fu, J. G. Learned, D. Marfatia, S. Pakvasa, and T. J. Weiler, Glashow resonance as a window into cosmic neutrino sources, Phys. Rev. D **90**, 121301 (2014).

[7] L. Wolfenstein, Neutrino oscillations in matter, Phys. Rev. D **17**, 2369 (1978).

[8] S. P. Mikheyev and A. Yu. Smirnov, Resonance amplification of oscillations in matter and spectroscopy of solar neutrinos, Yad. Fiz. **42**, 1441 (1985) [Sov. J. Nucl. Phys. **42**, 913 (1985)].

[9] V. Barger, K. Whisnant, S. Pakvasa, and R. J. N. Phillips, Matter effects on three-neutrino oscillations, Phys. Rev. D **22**, 2718 (1980).

[10] A. M. Dziewonski and D. L. Anderson, Preliminary reference Earth model, Phys. Earth Planet. Inter. **25**, 297 (1981).

[11] B. Feldstein, A. Kusenko, S. Matsumoto, and T. T. Yanagida, Neutrinos at IceCube from heavy decaying dark matter, Phys. Rev. D **88**, 015004 (2013).

[12] A. Esmaili and P. D. Serpico, Are IceCube neutrinos unveiling PeV-scale decaying dark matter? J. Cosmol. Astropart. Phys. 11 (2013) 054.

[13] K. Ioka and K. Murase, IceCube PeV–EeV neutrinos and secret interactions of neutrinos, Prog. Theor. Exp. Phys. **2014**, 061E01 (2014).

[14] K. C. Y. Ng and J. F. Beacom, Cosmic neutrino cascades from secret neutrino interactions, Phys. Rev. D **90**, 065035 (2014); **90**, 089904(E) (2014).

[15] S. Hümmer, M. Rüger, F. Spanier, and W. Winter, Simplified models for photohadronic interactions in cosmic accelerators, Astrophys. J. **721**, 630 (2010).

[16] A. Mücke, R. Engel, J. P. Rachen, R. J. Protheroe, and T. Stanev, SOPHIA: Monte Carlo simulations of photohadronic processes in astrophysics, Comput. Phys. Commun. **124**, 290 (2000).

[17] M. G. Aartsen, R. Abbasi, M. Ackermann, J. Adams, J. A. Aguilar, M. Ahlers, M. Ahrens, C. Alispach, N. M. Amin, K. Andeen, *et al.* (IceCube Collaboration), Detection of a particle shower at the Glashow resonance with IceCube, Nature (London) **591**, 220 (2021).

[18] K. Murase, D. Guetta, and M. Ahlers, Hidden cosmic-ray accelerators as an origin of TeV-PeV cosmic neutrinos, Phys. Rev. Lett. **116**, 071101 (2016).

[19] K. Murase, S. S. Kimura, and P. Mészáros, Hidden cores of active galactic nuclei as the origin of medium-energy neutrinos: Critical tests with the MeV gamma-ray connection, Phys. Rev. Lett. **125**, 011101 (2020).

[20] R. Abbasi, M. Ackermann, J. Adams, J. A. Aguilar, M. Ahlers, M. Ahrens, J. M. Alameddine, C. Alispach, A. A. Alves Jr., N. M. Amin, *et al.* (IceCube Collaboration), Evidence for neutrino emission from the nearby active galaxy NGC 1068, Science **378**, 538 (2022).

[21] R. Abbasi, M. Ackermann, J. Adams, S. K. Agarwalla, J. A. Aguilar, M. Ahlers, J. M. Alameddine, S. Ali, N. M. Amin, K. Andeen, *et al.* (IceCube Collaboration), Improved measurements of the TeV–PeV extragalactic neutrino spectrum from joint analyses of IceCube tracks and cascades, Phys. Rev. D **113**, 062002 (2026), companion to Ref. [2].

[22] N. Aghanim, Y. Akrami, M. Ashdown, J. Aumont, C. Baccigalupi, M. Ballardini, A. J. Banday, R. B. Barreiro, N. Bartolo, S. Basak, *et al.* (Planck Collaboration), Planck 2018 results. VI. Cosmological parameters, Astron. Astrophys. **641**, A6 (2020).

[23] P. Madau and M. Dickinson, Cosmic star-formation history, Annu. Rev. Astron. Astrophys. **52**, 415 (2014).

[24] Y. Ueda, M. Akiyama, G. Hasinger, T. Miyaji, and M. G. Watson, Toward the standard population synthesis model of the X-ray background: Evolution of X-ray luminosity and absorption functions of active galactic nuclei, Astrophys. J. **786**, 104 (2014).

[25] K. Carloni, Y. Porto, C. A. Argüelles, P. S. B. Dev, and S. Jana, Signatures of quasi-Dirac neutrinos in diffuse high-energy astrophysical neutrino data, Phys. Rev. D **113**, 115004 (2026).

[26] V. B. Valera, D. F. G. Fiorillo, I. Esteban, and M. Bustamante, New limits on neutrino decay from high-energy astrophysical neutrinos, Phys. Rev. D **110**, 043004 (2024).

[27] A. Cooper-Sarkar, P. Mertsch, and S. Sarkar, The high energy neutrino cross-section in the Standard Model and its uncertainty, J. High Energy Phys. 08 (2011) 042.

[28] J. Stachurska (IceCube Collaboration), First double cascade tau neutrino candidates in IceCube and a new measurement of the flavor composition, Proc. Sci. **ICRC2019**, 1015 (2020); R. Abbasi, M. Ackermann, J. Adams, J. A. Aguilar, M. Ahlers, M. Ahrens, C. Alispach, A. A. Alves Jr., N. M. Amin, K. Andeen, *et al.* (IceCube Collaboration), Detection of astrophysical tau neutrino candidates in IceCube, Eur. Phys. J. C **82**, 1031 (2022).

[29] N. Lad, T. J. van Eeden, and M. Ackermann (IceCube Collaboration), Neutrino flavor composition using high energy starting events with IceCube, Proc. Sci. **ICRC2025**, 1198 (2025).

[30] R. Abbasi, M. Ackermann, J. Adams, S. K. Agarwalla, J. A. Aguilar, M. Ahlers, J. M. Alameddine, N. M. Amin, K. Andeen, G. Anton, *et al.* (IceCube Collaboration), Observation of seven astrophysical tau neutrino candidates with IceCube, Phys. Rev. Lett. **132**, 151001 (2024).

[31] S. Aiello, A. Albert, A. R. Alhebsi, M. Alshamsi, S. Alves Garre, A. Ambrosone, F. Ameli, M. Andre, M. Anghinolfi, L. Aphecetche, *et al.* (KM3NeT Collaboration), Observation of an ultra-high-energy cosmic neutrino with KM3NeT, Nature (London) **638**, 376 (2025); **640**, E3(E) (2025).

[32] R. Abbasi, M. Ackermann, J. Adams, S. K. Agarwalla, J. A. Aguilar, M. Ahlers, J. M. Alameddine, N. M. Amin, K. Andeen, C. Argüelles, *et al.* (IceCube Collaboration), Search for extremely-high-energy neutrinos and first constraints on the ultrahigh-energy cosmic-ray proton fraction with IceCube, Phys. Rev. Lett. **135**, 031001 (2025).

[33] O. Adriani, S. Aiello, A. Albert, A. R. Alhebsi, M. Alshamsi, S. Alves Garre, A. Ambrosone, F. Ameli, M. Andre, L. Aphecetche, *et al.* (KM3NeT Collaboration), Ultra-high-energy event KM3-230213A within the global neutrino landscape, Phys. Rev. X **15**, 031016 (2025).

[34] K. Kotera, D. Allard, and A. V. Olinto, Cosmogenic neutrinos: Parameter space and detectability from PeV to ZeV, J. Cosmol. Astropart. Phys. 10 (2010) 013.

[35] O. Adriani, S. Aiello, A. Albert, A. R. Alhebsi, M. Alshamsi, S. Alves Garre, A. Ambrosone, F. Ameli, M. Andre, L. Aphecetche, *et al.* (KM3NeT, MessMapp Group, Fermi-LAT, Owens Valley Radio Observatory 40-m Telescope Group, and SVOM Collaborations), Characterizing candidate blazar counterparts of the ultra-high-energy event KM3-230213A, Astron. J. **172**, 139 (2026).

[36] C. Yuan, L. Pfeiffer, W. Winter, J. M. S. Zaballa, S. Buson, F. Testagrossa, and A. Azzollini, An accretion flare interpretation for the ultra-high-energy neutrino event KM3-230213A, Astrophys. J. Lett. **998**, L1 (2026).

[37] R. Abbasi, M. Ackermann, J. Adams, S. K. Agarwalla, J. A. Aguilar, M. Ahlers, J. M. Alameddine, S. Ali, N. M. Amin, K. Andeen, *et al.* (IceCube Collaboration), Evidence for neutrino emission from X-ray-bright active galactic nuclei with IceCube, Astrophys. J. Lett. **1000**, L26 (2026).

[38] IceCube, Fermi-LAT, MAGIC, AGILE, ASAS-SN, HAWC, H.E.S.S., INTEGRAL, Kanata, Kiso, Kapteyn, Liverpool Telescope, Subaru, Swift/NuSTAR, VERITAS, and VLA/17B-403 Teams, Multimessenger observations of a flaring blazar coincident with high-energy neutrino IceCube-170922A, Science **361**, eaat1378 (2018).

[39] M. G. Aartsen, M. Ackermann, J. Adams, J. A. Aguilar, M. Ahlers, M. Ahrens, I. Al Samarai, D. Altmann, K. Andeen, T. Anderson, *et al.* (IceCube Collaboration), Neutrino emission from the direction of the blazar TXS 0506+056 prior to the IceCube-170922A alert, Science **361**, 147 (2018).

[40] A. Zegarelli, A. Franckowiak, G. Sommani, N. Valtonen-Mattila, and T. Yuan (IceCube Collaboration), IceCat-2: Updated IceCube event catalog of alert tracks, Proc. Sci. **ICRC2025**, 1224 (2025).

[41] R. Kuze, K. Ioka, K. Murase, S. S. Kimura, and K. Inayoshi, Little red dots as hidden neutrino sources, Phys. Rev. D **113**, 083048 (2026).